\PassOptionsToPackage{table}{xcolor}
\documentclass[manuscript,screen]{acmart}

\usepackage{multirow}
\usepackage{makecell}
\usepackage{array} 
\usepackage{pifont}
\usepackage{graphicx}
\usepackage{subcaption}
\usepackage{booktabs}
\usepackage{wrapfig}
\usepackage{caption}
\usepackage{adjustbox}
\usepackage{float} 
\usepackage{placeins}
\usepackage[table]{xcolor}
\usepackage{adjustbox} 
\usepackage[many]{tcolorbox} 
\usepackage{fontawesome5}
\usepackage{colortbl}
\usepackage{xspace}
\usepackage[dvipsnames]{xcolor}

\def\ie{\emph{i.e.,}\xspace}

\newcommand{\cl}[1]{#1}
\newenvironment{cp}[0]{\par}{\par}
\newcommand{\cc}[0]{}

\AtBeginDocument{%
  }
\newtcolorbox{boxK}{
    top=2.2pt,
    bottom=2.2pt,
    left=4.5pt,
    right=4.5pt,
    boxrule = 0pt,
    toprule = 0pt,
    enhanced,
}

\newtcolorbox{boxRQ}{
    top=1.2pt,
    bottom=1.2pt,
    left=3.5pt,
    right=3.5pt,
    boxrule = 0pt,
    toprule = 0pt, 
    enhanced,
}

\setcopyright{none}
\copyrightyear{}
\acmYear{2026}
\acmDOI{}
\acmJournal{TOSEM}

\begin{document}

\makeatletter
\AddToHook{cmd/maketitle/before}{%
  \let\paper@original@fnsymbol\@fnsymbol
  \def\@fnsymbol#1{%
    \ifcase#1\or *\or \textsection\or \textdagger
    \else\paper@original@fnsymbol{#1}\fi}}
\AddToHook{cmd/maketitle/after}{%
  \let\@fnsymbol\paper@original@fnsymbol}
\makeatother

\title{Exploring the Potential of Diffusion Large Language Models in Code Generation}

\author{Chengze Li}
\orcid{0009-0008-0431-0125}
\authornote{Both authors contributed equally to this research.}
\authornote{\cl{This work was conducted while Chengze Li was an intern at Tsinghua University.}}
\affiliation{%
  \institution{College of AI, Tsinghua University}
  \city{Beijing}
  \country{China}
}
\affiliation{%
  \institution{School of Computer Science, Nanjing University}
  \city{Nanjing}
  \country{China}
}
\email{231220004@smail.nju.edu.cn}

\author{Yitong Zhang}
\orcid{0009-0000-1138-4503}
\authornotemark[1]
\affiliation{%
  \institution{College of AI, Tsinghua University}
  \city{Beijing}
  \country{China}
}
\email{zhangyt26@mails.tsinghua.edu.cn}

\author{Jia Li}
\orcid{0000-0002-5579-8852}
\authornote{Jia Li is the corresponding author.}
\affiliation{%
  \institution{College of AI, Tsinghua University}
  \city{Beijing}
  \country{China}
}
\email{jia_li@mail.tsinghua.edu.cn}

\author{Liyi Cai}
\orcid{0009-0002-7848-6007}
\affiliation{%
  \institution{School of Computer Science, Peking University}
  \city{Beijing}
  \country{China}
}
\email{cailiyi@stu.pku.edu.cn}

\author{Ge Li}
\orcid{0000-0002-5828-0186}
\affiliation{%
  \institution{School of Computer Science, Peking University} 
  \city{Beijing}
  \country{China}
}
\email{lige@pku.edu.cn}

\renewcommand{\shortauthors}{C. Li, Y. Zhang, J. Li, L. Cai and G. Li}

\begin{abstract}
Large Language Models (LLMs) have revolutionized the landscape of code generation.
Existing LLMs mainly employ autoregressive generation, \ie generating code token-by-token from left to right. However, the underlying autoregressive generation has two limitations in code generation. First, autoregressive LLMs typically generate only one token per step, showing low efficiency in practice. Second, programming is a non-sequential process involving back-and-forth editing, while autoregressive LLMs only employ the left-to-right generation order. These two intrinsic limitations hinder the further development of LLMs in code generation.

Recently, diffusion LLMs have emerged as a promising alternative. \cl{They provide two characteristics that are particularly relevant to code generation}: multi-token prediction (\ie generating multiple tokens at each step) and flexible generation order (\ie flexibly determining which positions to generate tokens). However, there is no systematic study exploring diffusion LLMs in code generation. To bridge the knowledge gap, we present the first empirical study of diffusion LLMs for code generation. Our study involves 7 representative diffusion LLMs and conducts experiments on a wide range of benchmarks. Based on the results, we summarize the following findings. \ding{182} \cl{Current diffusion LLMs show promising but uneven code generation ability, with open-source diffusion LLMs becoming competitive in several settings and closed-source diffusion LLMs showing stronger results.} For example, on MBPP+, the best-performing diffusion LLM achieves a pass@1 score of 79.1\%, compared with 73.3\% for \cl{the best evaluated autoregressive baseline}. \ding{183} Diffusion LLMs \cl{exhibit stronger length extrapolation behavior in our long-code-understanding experiments}. \ding{184} We explore factors impacting the effectiveness and efficiency of diffusion LLMs, and provide practical guidance. \ding{185} We discuss several promising future directions to improve diffusion LLMs on code generation. We open-source all source code, data, and results to facilitate future research.
\end{abstract}



\maketitle

\section{Introduction}
\label{sec:introduction}

During software development, there is a growing demand for automatic code generation due to the substantial labor required in manual programming~\cite{shin2021survey, li2023skcoder}. Code generation aims to automatically generate executable source code that fulfills natural language requirements. Over the past few years, Large Language Models (LLMs) have shown remarkable capabilities in code generation, achieving strong performance on widely used benchmarks and significantly enhancing developer productivity~\cite{jiang2024aixcoder, o1, li2025aixcoder, li2025structured, zhang2025focused, li2024acecoder}.

\begin{figure}[H]
    \centering
    \adjustbox{valign=t}{%
    \begin{minipage}[t]{0.53\linewidth}
        \centering
        \begin{subfigure}[t]{0.99\textwidth}
            \centering
            \includegraphics[width=\linewidth]{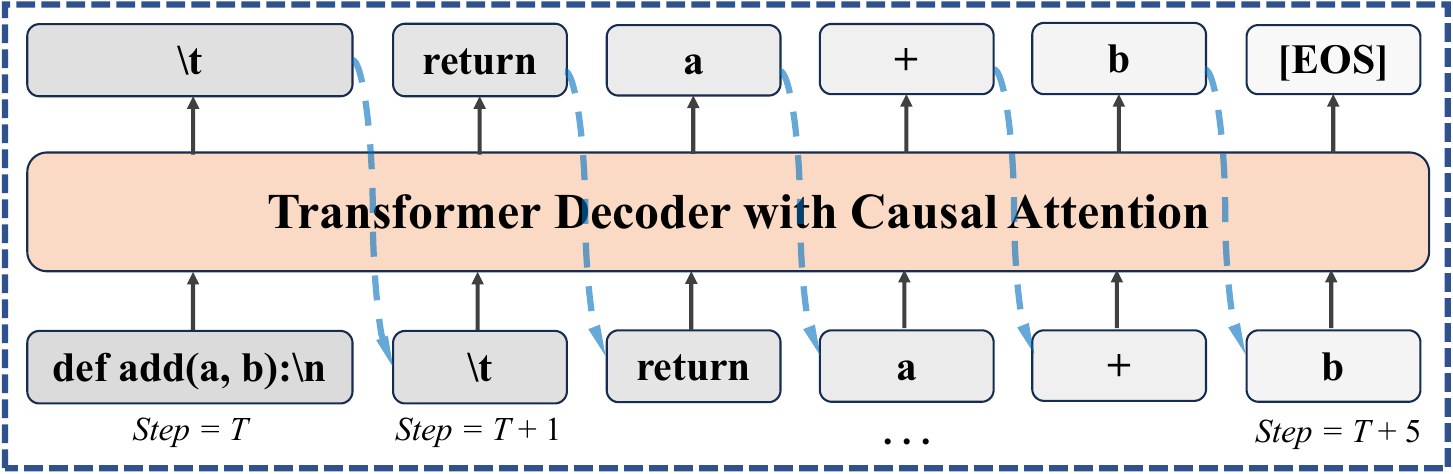}
            \caption{Autoregressive}
            \label{fig:modeling-AR}
        \end{subfigure}
        \begin{subfigure}[t]{0.99\textwidth}
            \centering
            \includegraphics[width=\linewidth]{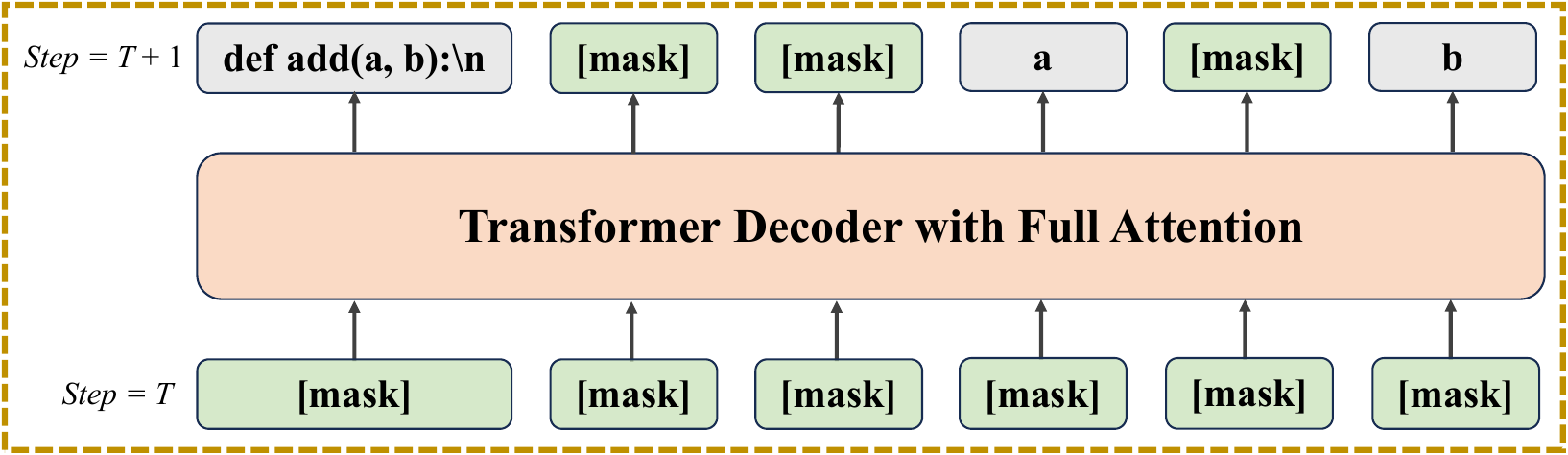}
            \caption{Diffusion}
            \label{fig:modeling-diffusion}
        \end{subfigure}
        \vspace{-0.08in}
        \caption{Generation paradigms of AR LLMs and diffusion LLMs.}
        \label{fig:modeling}
    \end{minipage}}%
    \hspace{0.15cm}
    \adjustbox{valign=t}{%
    \begin{minipage}[t]{0.38\linewidth}
        \centering
        \includegraphics[height=5.7cm]{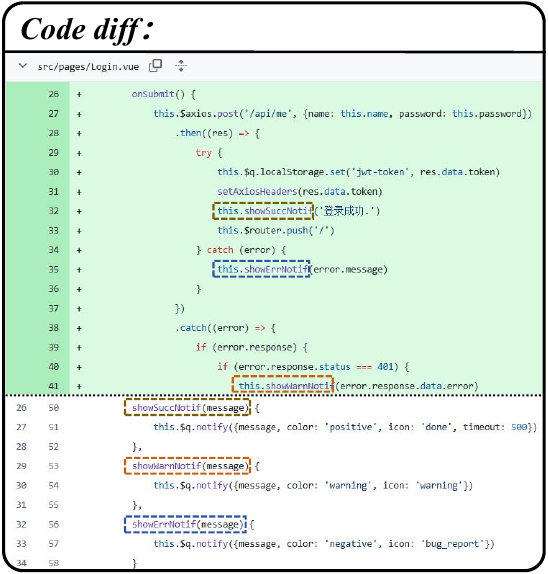}
        \caption{An example commit illustrating non-sequential programming.}
        \label{fig:code-diff}
    \end{minipage}}
    \Description{\cl{Comparison of autoregressive left-to-right generation, diffusion iterative generation, and a non-sequential programming example.}}
    \vspace{-0.22in}
\end{figure}

Building on this progress, a substantial body of recent work has explored the use of LLMs for code generation~\cite{jiang2024survey, zhang2026grammar, chen2026coact, cai2025ai}, with representative examples such as Qwen-Coder~\cite{qwencoder}, Seed-Coder~\cite{seed2025seed}, and DeepSeek-Coder~\cite{deepseekcoder, deepseek-coder-v2}. 
While these LLMs differ in various technical details, they are all built upon the autoregressive paradigm, where the model generates tokens strictly from left to right, as illustrated in Figure~\ref{fig:modeling-AR}.
However, there is an open question of \textbf{whether the mainstream autoregressive generation paradigm is always the best choice for code generation}. In practice, current \cl{autoregressive} LLMs (AR LLMs) reveal two \cl{important limitations that may restrict their use in some code generation scenarios}.
\cl{\ding{182} \textbf{Next-Token Prediction $\rightarrow$ Sequential-Decoding Cost.}} Most AR LLMs generate code strictly one token at a time, with each forward pass typically producing only a single token. \cl{Because each forward pass typically advances generation by one token, this sequential dependency can accumulate substantial computation and latency for long outputs.}
\cl{Such sequential latency can be particularly costly for repository-level generation~\cite{liao20243, wang2025teaching, bi2024iterative} and can reduce responsiveness in interactive coding assistants~\cite{liang2024large, sergeyuk2025using, wermelinger2023using}.}
\cl{\ding{183} \textbf{Left-to-Right Generation Order $\rightarrow$ Limited In-Sequence Revision.}}
\cl{AR LLMs decode in a strictly left-to-right manner, with each new token conditioned on the preceding context. This process can be less flexible than the non-sequential refinement common in programming workflows.}
For example, as shown in Figure~\ref{fig:code-diff}, a developer may first implement the body of the callee functions and later write the corresponding caller code, which directly contrasts with the rigid sequential nature of autoregressive decoding. \cl{Once an AR LLM has emitted an earlier portion of code, information introduced later cannot be used to revise that portion without an additional editing or regeneration pass.}
\cl{These constraints motivate the study of alternative paradigms that could complement autoregressive generation in some programming settings.}

\begin{cp}

Recently, a new generation paradigm of large language models, known as diffusion LLMs, has attracted increasing attention. Their generation process is illustrated in Figure~\ref{fig:modeling-diffusion}.  
Models such as LLaDA~\cite{llada, llada-v} and Dream~\cite{dream} have shown promising performance, suggesting that diffusion-based generation may serve as a viable alternative to the conventional autoregressive paradigm in some scenarios.
Compared with AR LLMs, diffusion LLMs exhibit two characteristics that are potentially relevant to code generation.  
\ding{182} \textbf{Multi-Token Prediction $\rightarrow$ Potential Efficiency Benefits.} Enabled by parallel decoding, diffusion LLMs can predict multiple tokens within a single forward pass. This property may reduce the number of decoding iterations needed to produce an output and thus offers a potential path toward more efficient generation.
Recent diffusion LLMs have reported encouraging inference efficiency while maintaining competitive performance on several benchmarks, suggesting their possible utility in latency-sensitive code generation scenarios~\cite{seeddiffusion, geminidiffusion}.
\ding{183} \textbf{Flexible Generation Order $\rightarrow$ Potential Alignment with Human Programming.}  
Rather than generating tokens strictly from left to right, diffusion LLMs can update different positions during the iterative denoising process. This flexibility may better accommodate the non-sequential refinement patterns often observed in human programming~\cite{diffucoder}. For example, a diffusion LLM may first generate callee functions before producing the corresponding caller functions, or prioritize parts of the code with richer contextual information before filling in those with less context.
\cl{However, systematic, code-generation-centered evaluations that jointly examine effectiveness, efficiency, inference settings, and practical limitations remain limited.}
\end{cp}

\cl{In this paper, we conduct one of the first comprehensive, code-generation-centered empirical studies of diffusion LLMs.} To address the existing gap and to guide future development, we propose four main Research Questions (RQs) together with their corresponding experiments. Building on these experiments, we provide a detailed analysis that uncovers several important findings. In the following, we introduce each research question in turn.

\vspace{-2pt} \begin{boxRQ}
\small \faIcon{search} \textit{\textbf{RQ1:}
How do diffusion LLMs compare with \cl{representative} AR LLMs in code generation tasks?
}\end{boxRQ} \vspace{-4.5pt}
\cl{To address this question, we evaluate 7 representative diffusion LLMs alongside 4 representative open-source AR baselines, treating the closed-source diffusion model as a separate reference point.}
\cl{Our results show that current diffusion LLMs exhibit promising but uneven code generation ability. Open-source diffusion LLMs have not yet consistently matched the strongest open-source AR baselines, but they are already competitive in several settings, and some evaluated diffusion models solve problems not solved by the selected AR models.} This complementary strength highlights their potential as an alternative modeling paradigm.

\begin{cp}
\vspace{-2pt} \begin{boxRQ}
\small \faIcon{search} \textit{\textbf{\cl{RQ2}:}
How well do diffusion LLMs perform on \cl{long-context code understanding}?
}\end{boxRQ} \vspace{-4.5pt}
\cl{Long-context code understanding is an important capability for repository-level software engineering tasks. In this RQ, we examine how diffusion LLMs behave when processing long code contexts. To this end, we conduct experiments on RepoQA~\cite{repoqa} and RepoBench-C~\cite{repobench}, evaluating three diffusion LLMs and a comparable AR baseline across a range of context lengths.}
\cl{Across the common 1k--8k input range, the evaluated diffusion LLMs are more robust to increasing input length than the comparable-scale AR baseline, although the comparison is not fully context-window matched.}
\end{cp}

\vspace{-2pt} \begin{boxRQ}
\small \faIcon{search} \textit{\textbf{\cl{RQ3}:}
What factors influence the effectiveness of code generation by diffusion LLMs? 
}\end{boxRQ} \vspace{-4.5pt}
In this RQ, we investigate how different settings influence the effectiveness of diffusion LLMs. We consider a range of key factors and evaluate several representative models on \cl{HumanEval, HumanEval-X~\cite{humaneval-multilingual}, MBPP~\cite{mbpp}, and LiveCodeBench~\cite{livecodebench}}.
\cl{Our analysis shows that the preferred generation length is model- and task-dependent: overly short canvases can prevent complete solutions, whereas longer fixed canvases increase decoding cost and can reduce effectiveness in some configurations.}

\vspace{-2pt} \begin{boxRQ}
\small \faIcon{search} \textit{\textbf{\cl{RQ4}:}
\cl{How do diffusion steps and generation length affect the efficiency of diffusion LLMs?}
}\end{boxRQ} \vspace{-4.5pt}
\cl{Parallel decoding offers diffusion LLMs a potential path to efficient generation. In this RQ, we systematically evaluate how diffusion steps and generation length affect efficiency on HumanEval, HumanEval-X, MBPP, and LiveCodeBench.}
The results show that both increasing the number of diffusion steps and extending the generation length reduce efficiency. \cl{We also compare AR decoding with and without KV cache, which provides a more careful view of the practical efficiency trade-offs.} These findings provide practical guidance for balancing efficiency and accuracy in real-world code generation scenarios.

Building on the above results and analyses, we outline several promising directions for future research on diffusion LLMs for code generation. These directions span multiple perspectives, such as applying diffusion models to a broader range of suitable coding tasks, integrating the generation process with structural characteristics of code, and exploring other avenues that further enhance their performance and practicality.

To summarize, the key contributions of this paper are as follows:  
\begin{itemize}
    \item \cl{We conduct one of the first comprehensive, code-generation-centered empirical studies of emerging diffusion LLMs, highlighting their characteristics compared with AR LLMs.}
    \item \cl{We systematically analyze selected inference factors that influence the effectiveness and efficiency of diffusion LLMs in code generation, identifying configuration-specific patterns and trade-offs across the evaluated settings.}
    \item Based on our study results and prior work, we propose multiple promising directions for future research aimed at enhancing the application of diffusion LLMs in code generation, \cl{while also discussing their current limitations}. 
\end{itemize}

\section{Background and Related Work}
\label{sec:Background}

\begin{cp}
\subsection{Code Generation}
\label{sec:code_generation}

Code generation is a long-standing task in software engineering that aims to automatically synthesize source code from natural language requirements~\cite{shin2021survey, li2023skcoder, jiang2024survey, li2025aixcoder}. 
With the rapid development of large language models, code generation has evolved from task-specific neural models to general-purpose and code-specialized LLMs that can support a wide range of programming scenarios~\cite{li2024acecoder, ouyang2025empirical, li2026papers, liu2026packmonitor}. 
Representative code LLMs, such as Qwen-Coder~\cite{qwencoder}, Seed-Coder~\cite{seed2025seed}, and DeepSeek-Coder~\cite{deepseekcoder, deepseek-coder-v2}, have demonstrated strong performance on standard code generation benchmarks and have become important components of modern coding assistants.

Existing studies on code generation typically evaluate models from multiple perspectives. 
Function-level benchmarks such as HumanEval~\cite{humaneval} and MBPP~\cite{mbpp} measure whether a model can synthesize correct programs from concise natural language specifications, while enhanced variants such as HumanEval+ and MBPP+ provide stronger test suites for more reliable functional correctness evaluation~\cite{evalplus}. 
Multilingual benchmarks such as HumanEval-X~\cite{humaneval-multilingual} further examine whether code generation ability generalizes across programming languages. 
More recent benchmarks, such as LiveCodeBench~\cite{livecodebench}, aim to reduce potential contamination by using continuously updated programming problems. 
Beyond function-level synthesis, repository-level benchmarks such as RepoQA~\cite{repoqa} and SWE-Bench~\cite{jimenez2024swebench} evaluate whether models can understand long code contexts and support more realistic software engineering tasks, such as generating code that correctly uses project-private libraries~\cite{zhang2026masterteachingllmsuse}.

A growing body of empirical research has investigated the practical strengths, limitations, and developer experience of LLM-based code generation~\cite{ouyang2025empirical, liang2024large, sergeyuk2025using, wermelinger2023using, jiang2024survey}. 
These studies provide useful insights for both researchers and practitioners, ranging from how developers interact with code assistants in daily programming workflows to how different model behaviors and usage settings affect the functional correctness, usability, and reliability of generated code. 
\cl{Broadly, existing studies examine code generation from several complementary perspectives, including benchmark-based functional correctness, multilingual generalization, repository-level context understanding, developer interaction, and deployment-oriented concerns such as latency and reliability.}
For example, prior empirical studies have examined the effectiveness of LLMs across different programming tasks~\cite{ouyang2025empirical, jiang2024survey}. 
Other studies have analyzed how developers interact with code assistants in practice and what challenges they encounter during daily programming workflows~\cite{liang2024large, sergeyuk2025using, wermelinger2023using}. 
These studies further show that model behavior, prompting strategies, and usage contexts can substantially influence the functional correctness, usability, and reliability of generated programs~\cite{liang2024large, ouyang2025empirical}.

However, most existing empirical studies and code generation models are built upon the autoregressive generation paradigm, where code is generated token by token from left to right. 
This paradigm has achieved substantial success and remains the dominant approach in current code LLMs. 
Nevertheless, relatively little is known about whether alternative generation paradigms can provide complementary benefits for code generation. 
This question is particularly relevant to code generation because practical coding systems place strong demands on inference efficiency, especially in interactive settings where developers expect timely feedback~\cite{liang2024large, sergeyuk2025using, wermelinger2023using}. 
Moreover, programming is not always a strictly left-to-right process: developers often edit, revise, and refine different parts of a program in a non-sequential manner. 
These characteristics make it valuable to examine whether alternative generation paradigms can complement the dominant autoregressive paradigm for code generation.
In this work, we focus on diffusion LLMs as one such emerging paradigm and empirically study their potential, limitations, and practical trade-offs for code generation.
\end{cp}

\subsection{Autoregressive Large Language Models}
\label{sec:arllm}

Autoregressive generation represents the predominant paradigm for large language models, commonly referred to as AR LLMs. It models a text sequence $Y = \{y_1, y_2, \ldots, y_L\}$ typically by producing one token at a time in a left-to-right manner. The generation of each token is conditioned on the sequence of previously generated tokens, a process formalized by the chain rule of probability:
\begin{equation}
\cc
p(Y \mid X) = \prod_{i=1}^{L} p(y_i \mid y_{<i}, X),
\label{eq:ar}
\end{equation}
\cl{where \( X \) denotes the input context and \( y_{<i} \) represents the tokens preceding position \( i \).}

\cl{Building on the success of general-purpose LLMs such as GPT-4~\cite{achiam2023gpt},} extensive research has applied autoregressive LLMs to code generation, yielding substantial gains in quality and functional correctness~\cite{li2024acecoder, ouyang2025empirical}. This progress has led to specialized state-of-the-art models fine-tuned for programming, such as Qwen-Coder~\cite{qwencoder} and DeepSeek-Coder~\cite{deepseekcoder}, which now serve as the core engines behind many modern coding assistants and agents~\cite{islam2025codesim, mowar2025codea11y}.

Despite their success, the autoregressive paradigm presents limitations that \cl{may be} misaligned with the practical demands of code generation. First, next-token prediction constrains efficiency and creates a latency bottleneck, which poses challenges for repository-level code generation and for real-time applications such as coding assistants and agents. Second, the strictly left-to-right generation process \cl{can be less flexible than} human programming practices in many cases, which typically involve non-sequential refinement. These \cl{potential mismatches} highlight the need to explore alternative generation paradigms that \cl{may complement autoregressive models in code generation}.

\subsection{Diffusion Large Language Models}
\label{sec:dllm}

Diffusion language models have recently become a focal point in AI research, offering an alternative to traditional autoregressive models~\cite{sahoo2024simple, yang2026improving, discreteVLA, yang2025mmada, yang2025difftester, li2026diffuguard}. They can be broadly categorized into continuous and discrete formulations. Empirical evidence suggests that discrete diffusion language models scale more effectively to large model sizes~\cite{gong2024scaling}, which has led to the development of diffusion large language models, such as LLaDA~\cite{llada}, Dream~\cite{dream}, and Gemini Diffusion~\cite{geminidiffusion}. Among these, most approaches typically adopt a masked discrete diffusion paradigm~\cite{sahoo2024simple}. To illustrate how diffusion LLMs work in practice, we present a representative example of the training and inference procedure. \textbf{During training}, an original sequence $Y_0=\{y_1^0,\ldots,y_L^0\}$ is corrupted into $Y_t$ along a continuous schedule $t\in[0,1]$ via independent masking, with a simple corruption process:
\begin{equation}
q(y_i^t \mid y_i^0)=
\begin{cases}
1-t & \text{if } y_i^t=y_i^0\\
t & \text{if } y_i^t=\texttt{[MASK]}
\end{cases}
\label{eq:diffusion}
\end{equation}
where \texttt{[MASK]} denotes a special token. The model is trained to reverse this corruption process by minimizing a cross-entropy loss computed only on masked positions. 
\textbf{During inference}, generation begins from a fully masked sequence with a predefined generation length. At each step, the model first predicts tokens for all masked positions. It then uses a specific remasking strategy to retain a subset of the predictions, while the remaining positions are remasked for further refinement. This iterative refinement continues until no masks remain, enabling parallel and flexible non-sequential updates across positions. 
\cl{In practice, several key factors may shape the behavior of diffusion LLMs: (1) \textit{Generation length}, which sets the number of positions in the initially masked generation canvas, although the semantic output may terminate earlier with \texttt{EOS}; (2) \textit{Diffusion steps}, which specifies the number of denoising iterations, while the decoding schedule determines how many predictions are retained in each iteration; (3) \textit{Remasking strategy}, which decides which predicted tokens are retained and which are masked again at each step; (4) \textit{Block diffusion}, which divides the sequence into blocks that are decoded sequentially from left to right, with each block processed using the standard diffusion procedure; and (5) \textit{Decoding temperature}, which controls the randomness of outputs, with higher values producing more diverse generations.}

Although diffusion LLMs were not originally designed for code generation, their ability to perform parallel and flexible generation \cl{may be useful for} iterative refinement workflows in programming, motivating a growing line of code-oriented diffusion models. DiffuCoder~\cite{diffucoder}, Mercury Coder~\cite{mercurycoder}, and Seed Diffusion Preview~\cite{seeddiffusion} represent notable advances in this direction, introducing techniques such as coupled GRPO for reinforcement learning and edit-based corruption to correct errors during inference. LLaDA 2.0~\cite{bie2025llada20scalingdiffusionlanguage} further demonstrates the scalability of diffusion LLMs by extending to the 100B scale, providing a strong foundation for \cl{studying their broader capabilities}. TreeDiff~\cite{zeng2025treediff} incorporates syntax awareness by masking abstract syntax tree subtrees during training, thereby improving syntactic correctness and structured reconstruction. In addition, recent work~\cite{mundler2025constrained,zhang2026lookahead} proposes constrained decoding for diffusion LLMs that enforces context-free grammars.

\begin{cp}
Prior code-oriented diffusion studies have introduced specialized models and techniques for generation, infilling, and repair~\cite{singh2023codefusion,singh2025diffusion,diffucoder,dreamon}, while block diffusion provides a continuum between autoregressive and full-sequence diffusion decoding~\cite{arriola2025block}.
These studies primarily focus on individual modeling or decoding techniques.
Concurrently with our work, Zhang et al.~\cite{zhang2025exploringpowerdiffusionlarge} present an empirical investigation of diffusion LLMs across the software development lifecycle, covering tasks such as code generation, defect detection, and program repair.
In contrast, our study centers on code generation and jointly examines cross-model effectiveness, inference-setting sensitivity, efficiency trade-offs, and long-context behavior.
Comprehensive evaluations that combine these dimensions across diverse diffusion LLM families and representative autoregressive baselines remain limited.
\end{cp}

\section{Study Design}
\label{sec:design}

\subsection{Overview and Research Questions}
\label{sec:rq}

\begin{figure}[H]
    \centering
    \definecolor{overviewpeach}{RGB}{238,217,203}
    \begin{tikzpicture}
        \node[inner sep=0] (overviewimage) {\includegraphics[width=0.91\textwidth]{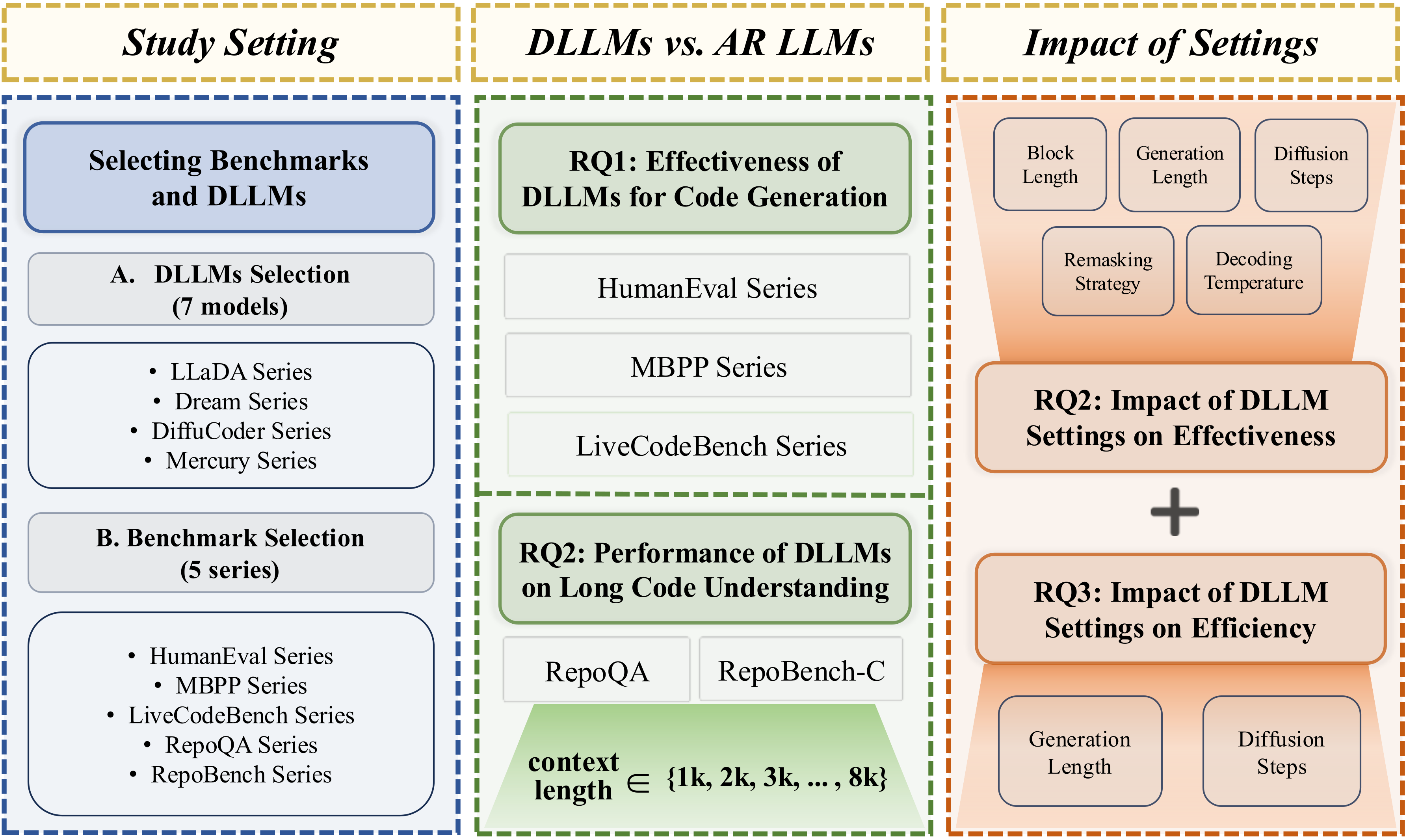}};
        \node[fill=overviewpeach, inner sep=1pt, minimum width=100pt, minimum height=27pt, align=center, font=\small\bfseries] at (135pt,0pt) {{RQ3: Impact of DLLM}\\{Settings on Effectiveness}};
        \node[fill=overviewpeach, inner sep=1pt, minimum width=100pt, minimum height=27pt, align=center, font=\small\bfseries] at (135pt,-52pt) {{RQ4: Impact of DLLM}\\{Settings on Efficiency}};
    \end{tikzpicture}
    \vspace{-5pt}
    \caption{Overview of our study. DLLM is used as an abbreviation for Diffusion LLM.}
    \label{fig:overview}
    \Description{{Overview linking model and benchmark selection to the four research questions.}}
    \vspace{-0.18in}
\end{figure}

The goal of this paper is to assess the potential of diffusion LLMs for code generation. We conduct a systematic evaluation across multiple widely used benchmarks and diverse experimental settings, comparing representative diffusion LLMs with representative AR baselines. Figure~\ref{fig:overview} presents an overview of our study. We next introduce the Research Questions (RQs) we aim to investigate. 

\textbf{RQ1 Effectiveness of Diffusion LLMs for Code Generation: How do diffusion LLMs compare with representative AR LLMs in code generation tasks?} In RQ1, we evaluate representative open-source and closed-source diffusion LLMs against representative open-source AR baselines across widely used benchmarks under \cl{empirically selected inference settings}. \cl{Because Mercury-Coder-Small is accessed only through a commercial API, we treat it as a closed-source reference point when interpreting the comparison.}

\cl{\textbf{RQ2 Performance of Diffusion LLMs on Long-Context Code Understanding: How well do diffusion LLMs perform on long-context code understanding?} Understanding and generating code at the repository level remains an open challenge for LLMs~\cite{li2025aixcoder, pan2024enhancing}. RQ2 compares three diffusion LLMs with a comparable-scale AR baseline over the same 1k--8k input range. Because the models do not all have the same nominal context window, we interpret the comparison as model-specific evidence about robustness to increasing input length rather than as a fully context-window-controlled comparison.}

\textbf{\cl{RQ3} Impact of Diffusion LLM Settings on Effectiveness: What factors impact the effectiveness of code generation by diffusion LLMs?} The usability and quality of the generated code are critical for real-world applications~\cite{liang2024large}. \cl{RQ3} focuses on how different inference settings influence the functional usability of outputs. We select a range of potentially influential factors and evaluate their performance through experiments across multiple settings.

\cl{\textbf{RQ4 Impact of Diffusion Steps and Generation Length on Efficiency: How do diffusion steps and generation length affect the efficiency of diffusion LLMs?} Parallel decoding offers diffusion LLMs a potential path to efficient generation~\cite{zhang2025survey}. RQ4 systematically evaluates these two factors, which directly determine the number of denoising passes and the sequence length processed per pass.}

\subsection{Studied LLMs}

In this section, we introduce the LLMs evaluated in this study. A summary of the selected LLMs is provided in Table~\ref{tab:setting_llm}.

\begin{table}[H]
\centering
\small
\caption{The summary of LLMs used in this study.}
\label{tab:setting_llm}
\adjustbox{max width=\linewidth}{
\begin{tabular}{@{}cccccc@{}}
\toprule
\textbf{Type} & \textbf{Domain} & \textbf{Model} & \textbf{Open-Source} & \textbf{Scale} & \textbf{Release Date} \\ \midrule
\multirow{7}{*}{Diffusion} & General & \textsc{LLaDA-8B-Instruct} & \ding{51} & 8B & 02/2025 \\ & General & \textsc{LLaDA-1.5} & \ding{51} & 8B & 05/2025 \\ & General & \textsc{Dream-v0-Instruct-7B} & \ding{51} & 7B & 04/2025 \\ & Code & \textsc{Dream-Coder-v0-Instruct-7B} & \ding{51} & 7B & 07/2025 \\ & Code & \textsc{DreamOn-v0-7B} & \ding{51} & 7B & 07/2025 \\ & Code & \textsc{DiffuCoder-7B-cpGRPO} & \ding{51} & 7B & 06/2025 \\ & Code & \textsc{Mercury-Coder-Small} & \ding{55} & -- & 02/2025 \\ \midrule
\multirow{5}{*}{Autoregressive} & General & \textsc{Qwen3-8B} & \ding{51} & 8B & 04/2025 \\ & General & \cl{\textsc{Llama-2-7B-chat-hf}} & \ding{51} & 7B & 07/2023 \\ & Code & \textsc{Seed-Coder-8B-Instruct} & \ding{51} & 8B & 05/2025 \\ & Code & \textsc{DeepSeek-Coder-6.7B-Instruct} & \ding{51} & 6.7B & \cl{11/2023} \\ & Code & \textsc{CodeLlama-7B-Instruct-hf} & \ding{51} & 7B & \cl{08/2023} \\
\bottomrule
\end{tabular}%
}
\end{table}

\subsubsection{Diffusion LLMs}
\label{sec:diffusionllms}
We evaluate 7 models drawn from 4 diffusion LLM families.

\textbf{LLaDA.} The LLaDA models are open-source diffusion LLMs trained entirely from scratch. \textsc{LLaDA-8B-Instruct}~\cite{llada} is the first open-source diffusion large language model in the LLaDA series. \cl{\textsc{LLaDA-1.5}}~\cite{llada15} extends this model by incorporating Variance-Reduced  Preference Optimization (VRPO) during training. We include both \textsc{LLaDA-8B-Instruct} and \textsc{LLaDA-1.5} in our experiments.

\textbf{Dream.} The Dream family is another line of open-source diffusion LLMs. \textsc{Dream-v0-Instruct-7B}~\cite{dream} is initialized from an autoregressive base model and provides the foundation for subsequent variants. \textsc{Dream-Coder-v0-Instruct-7B}~\cite{dreamcoder} targets code generation and adaptively selects its decoding style based on the task. \textsc{DreamOn-v0-7B}~\cl{\cite{dreamon}} explores dynamic expansion and contraction of mask tokens during inference. We include \textsc{Dream-v0-Instruct-7B}, \textsc{Dream-Coder-v0-Instruct-7B}, and \textsc{DreamOn-v0-7B} in our experiments.

\textbf{DiffuCoder.} \textsc{DiffuCoder-7B-cpGRPO}~\cite{diffucoder} is post-trained with coupled-GRPO on 21K code examples, yielding substantial gains in code generation performance. We include \textsc{DiffuCoder-7B-cpGRPO} in our experiments.

\textbf{Mercury.} Developed by Inception Labs, \textsc{Mercury-Coder-Small}~\cite{mercurycoder} is the first commercial-scale diffusion large language model that demonstrates strong performance on key coding benchmarks. We include \textsc{Mercury-Coder-Small} in our experiments \cl{as a closed-source reference point}.

\subsubsection{Autoregressive LLMs}
\cl{For comparison with diffusion LLMs, we select the following representative open-source AR LLMs at broadly similar parameter scales.}

\textbf{Qwen.} Developed by the Qwen team, \textsc{Qwen3-8B}~\cite{yang2025qwen3} \cl{is an 8B model in the Qwen3 series with strong code-generation performance}. We include \textsc{Qwen3-8B} in our experiments.

\textbf{Seed-Coder.} Developed by the Seed team, \textsc{Seed-Coder-8B-Instruct}~\cite{seed2025seed} attains leading results among open-source models of comparable size across diverse coding tasks. We include \textsc{Seed-Coder-8B-Instruct} in our experiments.

\textbf{DeepSeek-Coder.} Developed by DeepSeek, the DeepSeek Coder family consists of code models trained from scratch~\cite{deepseekcoder}. We include \textsc{DeepSeek-Coder-6.7B-Instruct} in our experiments.

\textbf{Code Llama.} Developed by Meta, Code Llama provides \cl{a widely used open-source reference point} for code. We include \textsc{CodeLlama-7B-Instruct-hf}~\cite{codellama} in our experiments.

\cl{Additionally, in RQ2, we select \textsc{Llama-2-7B-chat-hf}~\cite{touvron2023llama} as a comparable-scale AR baseline and evaluate all four models over the same 1k--8k input range. The public model documentation and configurations do not report a common nominal context window for all four systems, so this comparison controls the evaluated input lengths and approximate parameter scale, but not native context capacity or training differences.}

\subsection{Benchmark Selection}
We evaluate diffusion LLMs and AR LLMs on a suite of widely-used code generation benchmarks that collectively cover (i) function-level synthesis with executable tests, (ii) cross-language generalization across mainstream programming languages, (iii) realistic and continuously updated problem distributions, and (iv) repository-level long-context understanding. Below we describe each benchmark and explain why it is included.

\begin{cp}
\textit{HumanEval and HumanEval+.}
HumanEval~\cite{humaneval} is a Python function-level code generation benchmark containing 164 manually curated programming tasks. Each task provides a natural-language description, a function signature, and unit tests for functional verification. 
We include HumanEval because it is the most widely adopted function-level benchmark, enabling direct comparability with prior work.
We also report results on HumanEval+~\cite{evalplus}, an extended variant with more rigorous and diverse test cases, which offers a stricter assessment of functional correctness and reduces the chance of overfitting to weak test suites.

\textit{MBPP and MBPP+.}
MBPP~\cite{mbpp} contains approximately 974 Python programming problems, each with an English description, a reference solution, and a small set of test cases.
We include MBPP to complement HumanEval with a larger set of generally simpler, library-usage-heavy tasks, providing broader coverage of everyday programming patterns.
Similarly, MBPP+~\cite{evalplus} strengthens MBPP with enhanced tests, allowing us to evaluate robustness under stronger oracles.

\textit{HumanEval-X}
To assess cross-language generalization beyond Python, we additionally evaluate on HumanEval-X~\cite{humaneval-multilingual}, covering five languages: Python, C++, Java, Go, and JavaScript.
HumanEval-X preserves the HumanEval-style task format (natural-language requirement + function signature + tests) while varying the programming language, making it a controlled setting for comparing multilingual code generation capability across model families.
\end{cp}

\textit{LiveCodeBench.}
To evaluate performance in more realistic and continuously evolving coding scenarios, we adopt LiveCodeBench~\cite{livecodebench}, a dynamic benchmark that regularly curates problems from contemporary competitive programming platforms (e.g., LeetCode and CodeForces).
We report pass@1 on the LiveCodeBench 2410--2504 split, covering problems released between October 2024 and April 2025. We include LiveCodeBench because it better reflects distribution shift and recency effects than static benchmarks, providing an additional check on external validity.

\textit{RepoQA}
For long-context, repository-level code understanding, we use RepoQA~\cite{repoqa}. In particular, we focus on its Searching Needle Function task, where the model must retrieve a specific function from a repository given only a natural-language description. The task contains 500 queries spanning 50 repositories across five programming languages, making it suitable for studying repository-scale retrieval and comprehension under long inputs.
We include RepoQA to directly match the long-context requirement in \cl{RQ2} and to avoid confounding factors introduced by end-to-end tooling or multi-step execution.

\begin{cp}
\textit{RepoBench-C.}
To further evaluate repository-level completion beyond retrieval, we adopt RepoBench-C~\cite{repobench}, a benchmark for real-world multi-file code completion.
Given in-file context and cross-file context collected from other files in the same repository, RepoBench-C requires models to predict the next line of code.
It supports both Python and Java, and its updated v1.1 version categorizes test instances into multiple prompt-length levels, including 2k, 4k, 8k, 12k, and 16k, which allows us to evaluate model behavior under increasingly long repository-level contexts.
We include RepoBench-C because it complements RepoQA by shifting the evaluation from natural-language-driven function retrieval to cross-file next-line code completion, thereby providing an additional view of whether diffusion LLMs can effectively exploit repository-level context in realistic programming scenarios.

\textit{Benchmarks used for each research question.}
For RQ1, we report results on HumanEval/HumanEval+, MBPP/MBPP+, HumanEval-X, and LiveCodeBench, which jointly cover standard function-level synthesis, stronger test oracles, cross-language generalization, and realistic evolving problem distributions.
For RQ2, which focuses on long-context code understanding, we use RepoQA and RepoBench-C. RepoQA evaluates whether models can locate target functions from repository-level contexts, while RepoBench-C assesses their ability to exploit cross-file context for repository-level code completion.
\cl{For RQ3 and RQ4, we use HumanEval, HumanEval-X, MBPP, and LiveCodeBench. These benchmarks provide executable test-based evaluation across Python, multilingual, general-purpose, and more recently collected coding tasks, enabling systematic analysis of how different inference settings affect the effectiveness and efficiency of diffusion LLMs. For setting sweeps, the exact model--benchmark configurations are reported in the corresponding tables, and we keep the interpretation aligned with those configurations.}
\end{cp}
    
\subsection{Evaluation Metrics}
We adopt \cl{seven} metrics that are consistent with our research questions and are widely used in prior studies.  

\textbf{pass@k.} 
For HumanEval, MBPP, LiveCodeBench, and their variants, we evaluate functional correctness using pass@k, a widely adopted metric that measures the percentage of problems for which at least one of the $k$ generated solutions passes all provided test cases. 
As pass@k can be sensitive to sampling stochasticity introduced by temperature-based decoding, we repeat experiments that are potentially influenced by decoding randomness five times with different random seeds and report \cl{the mean and standard deviation in the form mean (standard deviation)} to \cl{characterize variability across runs}.

\textbf{Retrieval Accuracy.} For RepoQA, we follow the metric defined in the original work. Given a repository and a natural language description, the model outputs a target function. A prediction is counted as correct only if the generated function is the most similar to the ground-truth needle function among all functions in the repository under BLEU, and the BLEU score exceeds a fixed threshold. Retrieval accuracy is the proportion of queries that meet both conditions.

\begin{cp}
\textbf{Exact Match (EM).}
For RepoBench-C, we use Exact Match (EM) to evaluate next-line code completion.
EM measures whether the generated line exactly matches the ground-truth line.

\textbf{Edit Similarity (ES).}
For RepoBench-C, we also report Edit Similarity (ES), which measures the string-level similarity between the generated line and the ground-truth line.
Compared with EM, ES provides a softer measure of completion quality by assigning partial credit to predictions that are close to the reference but not exactly identical.
\end{cp}

\cl{\textbf{FLOPs per Token.}} To assess computational efficiency, we report the average floating point operations required to generate each output token. This metric reflects the computational cost of the inference process, providing a hardware-agnostic measure that complements throughput by focusing on efficiency rather than latency.

\textbf{Throughput.} To assess time efficiency, we report decoding throughput in tokens per second, calculated as the number of output tokens divided by the total decoding time. Decoding time is measured in wall-clock seconds from the first decoding call to the production of the final output.

\cl{For fixed-length diffusion decoding, the token count corresponds to positions in the decoded canvas. We therefore interpret throughput together with average decoding time, which provides the more direct end-to-end comparison across generation paradigms.}

\textbf{Average Decoding Time.}
To complement throughput, we report the average decoding time per problem, defined as the total decoding time divided by the number of evaluated problems. This metric reflects end-to-end latency at the task level.

\subsection{Implementation}
For closed-source LLMs, we access the models through their official APIs. 
For open-source models, we load the weights from their official repositories on \cl{Hugging Face}.
We adopt a zero-shot prompting strategy, where each model is provided only with the raw benchmark input as the prompt.

\cl{All efficiency experiments are conducted on a server equipped with eight NVIDIA A100-40GB GPUs.}

\begin{cp}
For efficiency evaluation, we report AR LLM results under two inference settings: with KV cache disabled and with KV cache enabled.
For diffusion LLMs, we do not apply KV cache or other decoding acceleration techniques, for two reasons.
First, the standard KV cache mechanism is fundamentally designed for autoregressive decoding.
It relies on causal attention and a strictly left-to-right generation order, under which previously generated tokens remain fixed and their key-value states can be safely reused in subsequent decoding steps.
In contrast, diffusion LLMs typically use bidirectional attention and refine tokens at arbitrary positions through iterative denoising.
Since the representation of each position may depend on the current states of many other positions, updating part of the sequence can invalidate previously computed key-value states.
As a result, the standard AR-style KV cache cannot be directly applied to diffusion decoding in its original form.
Second, although several recent studies have explored approximate, block-wise, or diffusion-specific KV caching mechanisms~\cite{jiang2025d,fastdllm, liu2025dllm, wang2025diffusion}, these techniques are still at an early stage.
They are often tied to specific model architectures or decoding procedures and have not yet been uniformly adopted across diffusion LLM families.
Therefore, to ensure a consistent and reproducible comparison, we evaluate all diffusion LLMs using their default decoding procedures without applying KV cache.
Meanwhile, we report AR LLM efficiency both with and without KV cache, allowing readers to interpret the results under both an unaccelerated setting and a more practical AR deployment setting.
\end{cp}

\section{Study Results}
\label{sec:results}

\subsection{RQ1: Effectiveness of Diffusion LLMs for Code Generation}
\label{sec:rq1}
\textbf{Motivation.}
\cl{In this section, we assess how current diffusion LLMs compare with representative AR baselines on code-generation benchmarks.}

\vspace{4pt} \noindent
\textbf{Setting.} 
\cl{For RQ1, our goal is to evaluate each model under empirically selected, model-specific inference configurations.
We therefore select model-specific inference settings based on the empirical findings from RQ3, where we systematically analyze the impact of different decoding and sampling strategies.
This protocol allows each architecture to use the decoding controls it supports; \cl{the reported results therefore characterize the evaluated systems under the selected configurations rather than a controlled effect of architecture alone}.
Using these selected settings, we evaluate all models on HumanEval, HumanEval+, HumanEval-X, MBPP, MBPP+, and LiveCodeBench.
\cl{For the principal cross-paradigm comparison, we compare open-source models only.}
Closed-source diffusion LLMs are reported separately as reference points, but are not used to support direct conclusions about the relative performance of the two paradigms.}
We repeat each evaluation five times and report the average and standard deviation.

\vspace{4pt} \noindent
\textbf{Results and Analyses.} 
The results are shown in Tables~\ref{tab:rq1_acc_main} and~\ref{tab:rq1_acc_multi}.

\begin{table}[!t]
\centering
\small
\caption{\cl{Effectiveness of each model for code generation on HumanEval / HumanEval+, MBPP / MBPP+, and LiveCodeBench. Values report mean (standard deviation) over five runs.}}
\vspace{-0.12in}
\label{tab:rq1_acc_main}
\adjustbox{max width=\linewidth}{
\begin{tabular}{@{}ccccccc@{}}
\toprule
\textbf{Type} & \textbf{Model}
& \textbf{HumanEval}
& \textbf{HumanEval+}
& \textbf{MBPP}
& \textbf{MBPP+}
& \textbf{LiveCodeBench} \\ \midrule
\multirow{7}{*}{Diffusion}
 & \textsc{LLaDA-8B-Instruct}            & 47.0(1.2)\% & 40.4(1.4)\% & 59.3(0.6)\% & 49.2(0.6)\% & 6.1(0.4)\%  \\
 & \textsc{LLaDA-1.5}                    & 43.9(1.8)\% & 38.7(1.7)\% & 57.1(1.0)\% & 47.3(0.6)\% & 6.5(0.5)\%  \\
 & \textsc{Dream-v0-Instruct-7B}         & 56.7(0.0)\% & 49.4(0.0)\% & 73.1(0.3)\% & 61.1(0.3)\% & 9.7(0.0)\%  \\
 & \textsc{Dream-Coder-v0-Instruct-7B}   & 76.2(0.0)\% & 70.7(0.0)\% & 82.3(0.0)\% & 70.1(0.0)\% & 19.6(0.0)\% \\
 & \textsc{DreamOn-v0-7B}                & 52.4(0.0)\% & 47.0(0.0)\% & 71.2(0.0)\% & 60.6(0.0)\% & 7.6(0.0)\%  \\
 & \textsc{DiffuCoder-7B-cpGRPO}         & 69.5(0.0)\% & 62.2(0.0)\% & 77.6(0.1)\% & 66.6(0.2)\% & 9.1(0.1)\%  \\
 & \textsc{Mercury-Coder-Small}          & 85.4(0.0)\% & 79.9(0.0)\% & {93.7(0.0)\%} & {79.1(0.0)\%} & 19.9(0.0)\% \\
\midrule
\multirow{4}{*}{Autoregressive}
 & \textsc{Qwen3-8B}                     & 85.1(0.8)\% & 78.8(1.4)\% & 84.3(0.7)\% & 73.3(0.5)\% & 24.9(1.0)\% \\
 & \textsc{Seed-Coder-8B-Instruct}       & {86.5(0.5)\%} & {80.4(0.6)\%} & 85.9(0.4)\% & 73.1(0.2)\% & 23.6(1.0)\% \\
 & \textsc{DeepSeek-Coder-6.7B-Instruct} & 77.4(0.0)\% & 70.7(0.0)\% & 79.4(0.0)\% & 68.8(0.0)\% & 12.6(0.0)\% \\
 & \textsc{CodeLlama-7B-Instruct-hf}     & 41.5(1.6)\% & 36.2(0.8)\% & 57.6(0.5)\% & 49.6(0.3)\% & 9.2(0.6)\%  \\
\bottomrule
\end{tabular}
}
\end{table}

\begin{cp}
\textit{\textbf{Comparison Among Diffusion LLMs.}} 

\cl{\ding{182} \textbf{The evaluated closed-source diffusion LLM shows a noticeable advantage over current open-source diffusion LLMs.}}
Among the evaluated diffusion models, \textsc{Mercury-Coder-Small} achieves the best overall performance across most benchmarks.
For example, it obtains 85.4\% on HumanEval, 93.7\% on MBPP, and 79.1\% on MBPP+, outperforming all open-source diffusion LLMs in these settings.
It also achieves strong results on HumanEval-X, with 87.2\% on C++, 91.5\% on Java, 81.7\% on Go, and 92.1\% on JavaScript.
\cl{Because the scale, training data, and post-training pipeline of \textsc{Mercury-Coder-Small} are not publicly documented, this gap cannot be attributed to any single factor. It nevertheless shows that the strongest evaluated closed-source diffusion system remains ahead of current open-source diffusion LLMs in these settings.}

\cl{\ding{183} \textbf{Code-oriented diffusion LLMs tend to perform better than general-purpose diffusion LLMs in our evaluation.}}
A clearer pattern among open-source diffusion LLMs is that models adapted specifically for code generation tend to outperform more general-purpose diffusion LLMs.
For example, \textsc{Dream-Coder-v0-Instruct-7B}, a code-oriented variant in the Dream family, improves over \textsc{Dream-v0-Instruct-7B} by 19.5 percentage points on HumanEval, 21.3 points on HumanEval+, 9.2 points on MBPP, 9.0 points on MBPP+, and 9.9 points on LiveCodeBench.
Similarly, \textsc{DiffuCoder-7B-cpGRPO}, which is post-trained on code examples with coupled-GRPO, achieves stronger overall performance than general-purpose diffusion LLMs such as \textsc{LLaDA-8B-Instruct} and \textsc{LLaDA-1.5}.
\cl{These results associate code-specific training and post-training with stronger code-generation performance in the evaluated models, without isolating their individual causal effects.}

\begin{table}[!t]
\centering
\small
\caption{\cl{Effectiveness of each model for code generation on HumanEval-X. Values report mean (standard deviation) over five runs.}}
\vspace{-0.12in}
\label{tab:rq1_acc_multi}
\adjustbox{max width=\linewidth}{
\begin{tabular}{@{}ccccccc@{}}
\toprule
\textbf{Type} & \textbf{Model}
& \textbf{Python}
& \textbf{C++}
& \textbf{Java}
& \textbf{Go}
& \textbf{JavaScript} \\ \midrule
\multirow{7}{*}{Diffusion}
 & \textsc{LLaDA-8B-Instruct}            & 47.0(1.2)\% & 26.8(0.9)\% & 32.9(0.7)\% & 22.0(1.0)\% & 39.3(1.5)\% \\
 & \textsc{LLaDA-1.5}                    & 43.9(1.8)\% & 28.9(1.8)\% & 31.3(0.6)\% & 24.8(1.2)\% & 42.4(1.1)\% \\
 & \textsc{Dream-v0-Instruct-7B}         & 56.7(0.0)\% & 40.2(0.0)\% & 45.7(0.0)\% & 32.0(0.3)\% & 58.7(0.2)\% \\
 & \textsc{Dream-Coder-v0-Instruct-7B}   & 76.2(0.0)\% & 60.4(0.0)\% & 61.6(0.0)\% & 57.9(0.0)\% & 73.8(0.0)\% \\
 & \textsc{DreamOn-v0-7B}                & 52.4(0.0)\% & 10.4(0.0)\% & 8.5(0.0)\%  & 4.3(0.0)\%  & 15.2(0.0)\% \\
 & \textsc{DiffuCoder-7B-cpGRPO}         & 69.5(0.0)\% & 43.9(0.0)\% & 48.4(0.5)\% & 40.2(0.0)\% & 52.0(0.2)\% \\
 & \textsc{Mercury-Coder-Small}          & 85.4(0.0)\% & {87.2(0.0)\%} & {91.5(0.0)\%} & {81.7(0.0)\%} & {92.1(0.0)\%} \\
\midrule
\multirow{4}{*}{Autoregressive}
 & \textsc{Qwen3-8B}                     & 85.1(0.8)\% & 71.3(1.8)\% & 73.5(2.5)\% & 65.5(1.2)\% & 78.9(0.8)\% \\
 & \textsc{Seed-Coder-8B-Instruct}       & {86.5(0.5)\%} & 78.2(0.6)\% & 73.7(0.6)\% & 73.9(1.4)\% & 81.5(0.8)\% \\
 & \textsc{DeepSeek-Coder-6.7B-Instruct} & 77.4(0.0)\% & 64.0(0.0)\% & 68.3(0.0)\% & 33.5(0.0)\% & 70.1(0.0)\% \\
 & \textsc{CodeLlama-7B-Instruct-hf}     & 41.5(1.6)\% & 30.0(1.4)\% & 34.3(1.7)\% & 25.7(1.9)\% & 36.8(0.7)\% \\
\bottomrule
\end{tabular}
}
\end{table}

\ding{184} \textbf{The cross-language performance gap of diffusion LLMs is largely driven by a subset of weaker models.}
\cl{To better understand the cross-language generalization of individual diffusion LLMs, we report the drop from Python to the average of the other four languages for each model.}
\cl{Among open-source diffusion LLMs, \textsc{DreamOn-v0-7B} exhibits the most severe degradation, dropping from 52.4\% on Python to an average of only 9.6\% across the other languages (a 42.8 percentage-point drop), which substantially increases the family-level average cross-language drop.}
\cl{In contrast, \textsc{Dream-Coder-v0-Instruct-7B} shows a much more moderate drop from 76.2\% to 63.4\% (12.8 pp), which is comparable to the cross-language drop observed for AR models such as \textsc{Qwen3-8B} (85.1\% to 72.3\%, 12.8 pp) and \textsc{Seed-Coder-8B-Instruct} (86.5\% to 76.8\%, 9.7 pp).}
\cl{\textsc{DiffuCoder-7B-cpGRPO} drops from 69.5\% to 46.1\% (23.4 pp), while \textsc{LLaDA-8B-Instruct} and \textsc{LLaDA-1.5} show drops of 16.7 pp and 12.0 pp, respectively.}
\cl{Notably, \textsc{Mercury-Coder-Small} even improves on average across non-Python languages (85.4\% to 88.1\%), indicating strong multilingual code generation capability.}
\cl{Overall, these results show that cross-language degradation is not uniform across the evaluated diffusion LLMs and therefore should not be attributed to the generation paradigm alone.}
\cl{Some code-oriented diffusion LLMs demonstrate cross-language robustness comparable to selected AR baselines, whereas other evaluated models exhibit substantially larger gaps.}
\end{cp}

\begin{cp}
\textit{\textbf{Comparison Between Diffusion and AR LLMs.}}  

\ding{182} \textbf{Open-source diffusion LLMs have not yet consistently matched strong AR LLMs in overall code generation performance.}  
When focusing on open-source models, diffusion LLMs still show a performance gap compared with strong AR baselines on several benchmarks.
For example, on LiveCodeBench, the best open-source diffusion LLM, \textsc{Dream-Coder-v0-Instruct-7B}, achieves 19.6\%, while \textsc{Qwen3-8B} and \textsc{Seed-Coder-8B-Instruct} reach 24.9\% and 23.6\%, respectively.
Similarly, on HumanEval and HumanEval+, \textsc{Dream-Coder-v0-Instruct-7B} achieves 76.2\% and 70.7\%, which remain lower than \textsc{Qwen3-8B} and \textsc{Seed-Coder-8B-Instruct}.
These results indicate that current open-source diffusion LLMs have not yet fully matched mature AR LLMs in overall code generation effectiveness, especially on more challenging and recent coding benchmarks.

\ding{183} \textbf{Nevertheless, open-source diffusion LLMs already show competitive performance in several settings.}  
Although open-source diffusion LLMs still lag behind the strongest AR baselines in aggregate, some models achieve competitive results on standard function-level benchmarks.
For instance, \textsc{Dream-Coder-v0-Instruct-7B} obtains 76.2\% on HumanEval and 70.7\% on HumanEval+, approaching \textsc{DeepSeek-Coder-6.7B-Instruct}, which achieves 77.4\% and 70.7\% on the same benchmarks.
It also achieves 82.3\% on MBPP and 70.1\% on MBPP+, narrowing the gap with strong AR baselines such as \textsc{Qwen3-8B} and \textsc{Seed-Coder-8B-Instruct}.
Meanwhile, \textsc{DiffuCoder-7B-cpGRPO} reaches 69.5\% on HumanEval and 62.2\% on HumanEval+, outperforming \textsc{CodeLlama-7B-Instruct-hf} by 28.0 and 26.0 percentage points, respectively.
\cl{These results suggest that diffusion LLMs are not uniformly superior to AR LLMs, but some open-source diffusion LLMs are competitive with representative AR baselines in several evaluated settings.}
\end{cp}

\textit{\textbf{Complementarity Between Diffusion and AR LLMs.}}

\begin{figure}[t]
    \centering
    \adjustbox{valign=t}{%
        \includegraphics[height=5cm]{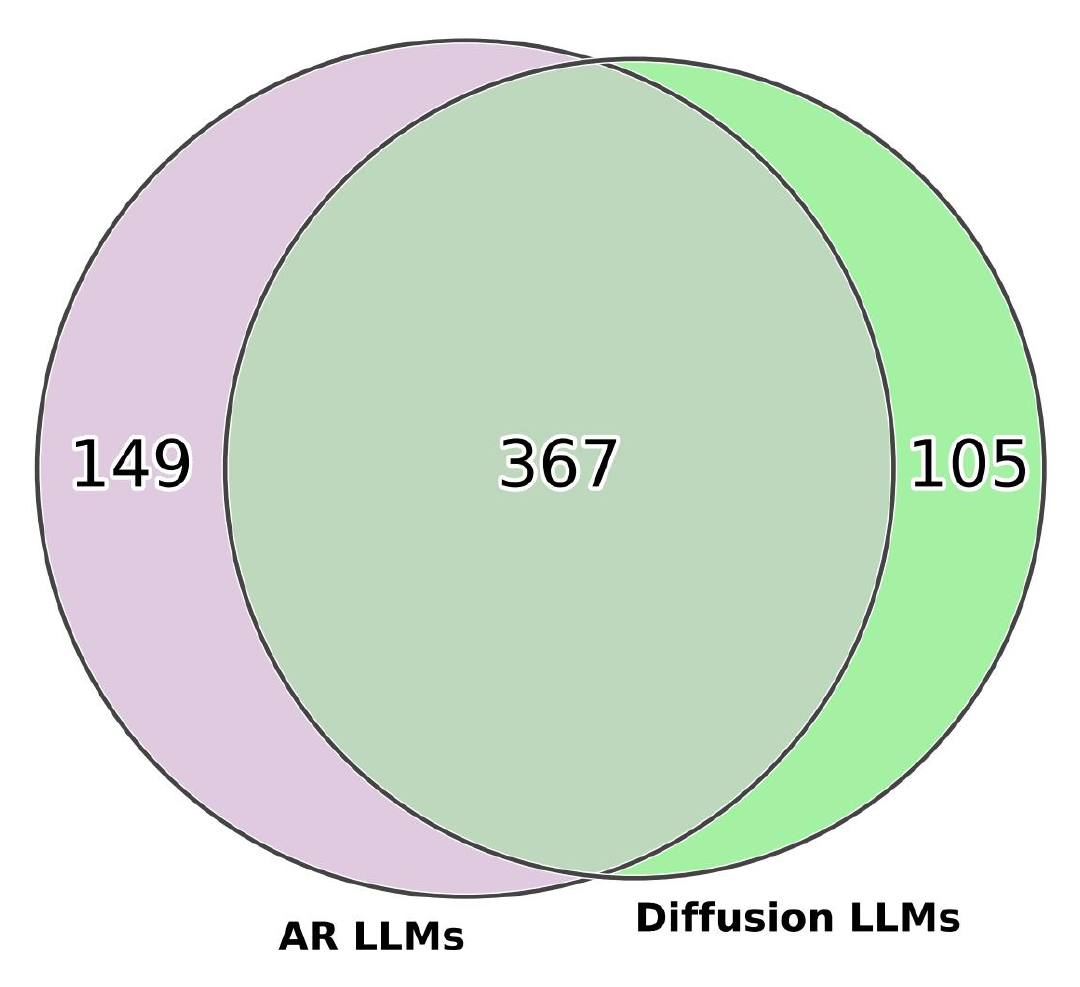}
    }
    \vspace{-8pt}
    \caption{\cl{Venn diagram of tasks solved by the selected diffusion and AR LLMs across HumanEval, MBPP, and the LiveCodeBench 2410--2504 split.}}
    \label{fig:rq1_veen}
    \Description{\cl{Venn diagram comparing tasks solved by the selected diffusion and autoregressive models.}}
\end{figure}

\begin{cp}
Although diffusion LLMs do not yet consistently outperform their AR counterparts in overall accuracy, we further investigate whether the selected model groups cover different problem instances. To this end, we select three representative diffusion LLMs: \textsc{Dream-Coder-v0-Instruct-7B}, \textsc{DiffuCoder-7B-cpGRPO}, and \textsc{Mercury-Coder-Small}, along with three strong AR LLMs: \textsc{Qwen3-8B}, \textsc{Seed-Coder-8B-Instruct}, and \textsc{DeepSeek-Coder-6.7B-Instruct}. We pool the evaluated problems from HumanEval, MBPP, and LiveCodeBench and, for each model group, identify the union of problems solved by at least one selected model. We then compare the two sets using a Venn diagram.
As shown in Figure~\ref{fig:rq1_veen}, 367 problems are solved by both model groups, while 149 are uniquely solved by the selected AR models and 105 are uniquely solved by the selected diffusion models. This non-overlap provides evidence of complementary problem coverage among the evaluated models, rather than establishing an inherent advantage of either paradigm.
\end{cp}

\begin{figure}[t]
    \centering
    \includegraphics[width=1.0\columnwidth]{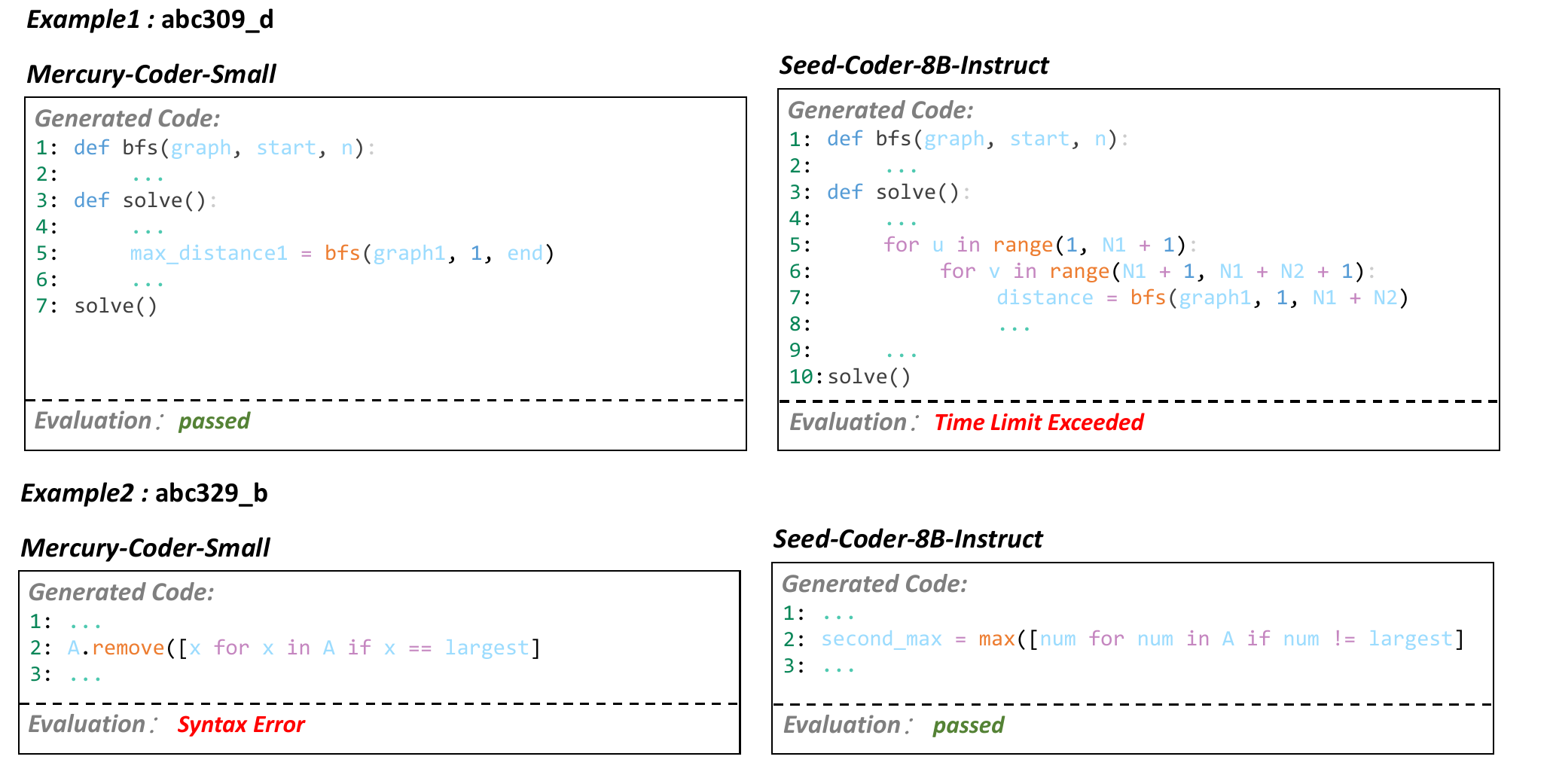}
    \caption{\cl{Examples illustrating complementary success and failure patterns of diffusion and AR LLMs. Ellipses indicate omitted portions of the generated code.}}
    \label{fig:complement_example}
    \Description{\cl{Two code-generation examples showing complementary successes and failures of diffusion and autoregressive models.}}
\end{figure}

\begin{cp}
To make this non-overlap concrete, we present two illustrative cases in Figure~\ref{fig:complement_example}, selected from the problems uniquely solved by one model group.
\ding{182} In the first case, the diffusion model produces an efficient solution with an auxiliary breadth-first-search function, whereas the AR model generates a nested-loop implementation that exceeds the time limit.
\ding{183} \cl{In the second case, the AR model's full output passes the benchmark tests, whereas the diffusion model's output fails with a syntax error; the figure uses ellipses to omit unshown portions of both outputs.}
These cases illustrate different observed success and failure modes; they do not by themselves establish that either behavior is caused solely by the underlying generation paradigm.
\end{cp}

\cl{\textit{\textbf{Performance on Repository-Level Tasks.}}}
\cl{While the above results focus on function-level code generation, we also evaluate the diffusion LLMs on SWE-Bench~\cite{jimenez2024swebench}, a widely used benchmark for repository-level bug fixing.}
\cl{We test all six open-source diffusion LLMs---\textsc{LLaDA-8B-Instruct}, \textsc{LLaDA-1.5}, \textsc{Dream-v0-Instruct-7B}, \textsc{Dream-Coder-v0-Instruct-7B}, \textsc{DreamOn-v0-7B}, and \textsc{DiffuCoder-7B-cpGRPO}---on the SWE-Bench Verified split.}
\cl{Under our zero-shot evaluation setting, all models achieve a resolved rate of 0\%.}
\cl{SWE-Bench requires multi-step reasoning across multiple files, understanding complex project structures, and generating precise patches, whereas the evaluated diffusion LLMs are primarily designed for function-level generation and do not provide an integrated repository-level agent workflow.}
\cl{We therefore treat this result as a preliminary boundary case for the evaluated models rather than a general conclusion about diffusion-based repository-level program repair.}

\begin{boxK}
\small \faIcon{pencil-alt} \textbf{Answer to RQ1:}
\cc
Current diffusion LLMs demonstrate promising but uneven code generation ability.
\cl{The evaluated closed-source diffusion system is the strongest diffusion model in our evaluation, while code-oriented open-source diffusion models tend to perform better than general-purpose diffusion models.}
Compared with open-source AR LLMs, open-source diffusion LLMs have not yet consistently matched the strongest AR baselines in overall effectiveness, particularly on more challenging benchmarks such as LiveCodeBench.
\cl{Nevertheless, they are already competitive on several standard function-level benchmarks and outperform some representative AR baselines in specific settings.}
\cl{Moreover, the selected diffusion and AR model groups solve partially different sets of problems, suggesting complementary problem coverage in the evaluated settings.}
\cl{Overall, diffusion LLMs are not uniformly superior to AR LLMs, but the evaluated systems show promising and partially complementary code-generation capabilities.}
\end{boxK}

\begin{cp}

\subsection{\cl{RQ2: Performance of Diffusion LLMs on Long-Context Code Understanding}}
\label{sec:rq4}

\textbf{Motivation.}
\cl{Code generation often requires LLMs to understand long code contexts, such as repository-level code generation~\cite{EvoCodeBench}. In this RQ, we evaluate the long-context code understanding ability of diffusion LLMs.}

\vspace{4pt} \noindent
\textbf{Setting.} 
To assess long-context behavior from complementary angles, we conduct experiments on two widely used repository-level benchmarks.
\cl{\textbf{RepoQA}~\cite{repoqa} formulates long-context code understanding as a retrieval task, where the model is asked to identify a target function in a repository given a natural-language description; we report retrieval accuracy as the metric.}
\textbf{RepoBench-C}~\cite{repobench} formulates it as a code completion task that requires the model to predict the next line of code given both in-file and cross-file context; we report Exact Match (EM) and Edit Similarity (ES) as metrics.
RepoQA varies the input context length from 1k to 8k tokens, while RepoBench-C reports results at 2k, 4k, and 8k.

\cl{We evaluate three representative diffusion LLMs---\textsc{LLaDA-1.5}, \textsc{DiffuCoder-7B-cpGRPO}, and \textsc{Dream-Coder-v0-Instruct-7B}---and one comparable-scale AR baseline, \textsc{Llama-2-7B-chat-hf}, over the same input-length range. Public model documentation and configurations do not report a common nominal context window for all four systems. We therefore use this experiment to compare robustness over the common 1k--8k range, while treating the results as model-specific rather than as a fully context-window-controlled comparison.}

\vspace{4pt} \noindent
\textbf{Results and Analyses.}
The complete results are reported in Table~\ref{tab:repoqa_context_perf} for RepoQA and Table~\ref{tab:repobenchc_context_perf} for RepoBench-C.
\cl{Both benchmarks are reported in full in the two tables, including every evaluated input length and metric.}

\begin{table}[H]
\centering
\small
\caption{\cl{Evaluation results on RepoQA. Each cell reports retrieval accuracy under the corresponding input context length; bold indicates the row-wise maximum.}}
\label{tab:repoqa_context_perf}
\adjustbox{max width=\linewidth}{
\begin{tabular}{l c c c c c c c c}
\toprule
\multirow{2}{*}{\textbf{Model}} & \multicolumn{8}{c}{\textbf{Input Context Length}} \\
\cmidrule(lr){2-9}
& 1k & 2k & 3k & 4k & 5k & 6k & 7k & 8k \\
\midrule
\textsc{Llama-2-7B-chat-hf} & \textbf{61.2\%} & 56.8\% & 46.2\% & 9.0\% & 2.3\% & 1.2\% & 1.2\% & 1.2\%\\

\textsc{DiffuCoder-7B-cpGRPO} & 37.2\% & 33.5\% & \textbf{40.2\%} & 32.5\% & 34.0\% & 37.0\% & 28.5\% & 28.5\%\\

\textsc{Dream-Coder-v0-Instruct-7B} & \textbf{44.0\%} & 32.8\% & 32.5\% & 28.5\% & 28.0\% & 19.5\% & 11.2\% & 8.3\%\\

\textsc{LLaDA-1.5} & \textbf{57.2\%} & 48.5\% & 33.2\% & 30.7\% & 27.5\% & 20.5\% & 17.5\% & 14.8\%\\

\bottomrule
\end{tabular}
}
\vspace{-0.04in}
\end{table}

\begin{table}[H]
\centering
\small
\caption{\cl{Evaluation results on RepoBench-C. Each cell reports Exact Match (EM) or Edit Similarity (ES) under the corresponding input context length; bold indicates the row-wise maximum for each metric.}}
\label{tab:repobenchc_context_perf}
\adjustbox{max width=\linewidth}{
\begin{tabular}{l c c c c c c}
\toprule
\multirow{2}{*}{\textbf{Model}} & \multicolumn{3}{c}{Exact Match} & \multicolumn{3}{c}{Edit Similarity} \\
\cmidrule(lr){2-4} \cmidrule(lr){5-7}
& 2k & 4k & 8k & 2k & 4k & 8k \\
\midrule
\textsc{Llama-2-7B-chat-hf} & \textbf{24.3\%} & 22.1\% & 3.4\% & \textbf{66.7\%} & 62.8\% & 19.2\%\\

\textsc{LLaDA-1.5} & 27.9\% & 27.8\% & \textbf{30.3\%} & 69.1\% & 68.2\% & \textbf{70.3\%}\\

\textsc{Dream-Coder-v0-Instruct-7B} & 30.6\% & 32.2\% & \textbf{33.3\%} & 72.3\% & 72.1\% & \textbf{73.3\%}\\

\textsc{DiffuCoder-7B-cpGRPO} & 27.9\% & 27.4\% & \textbf{29.4\%} & 64.5\% & 65.6\% & \textbf{66.5\%}\\

\bottomrule
\end{tabular}
}
\vspace{-0.04in}
\end{table}

\cl{\textbf{Across inputs up to 4k, the relative ordering depends on the benchmark and input length.}}
\cl{On RepoQA, \textsc{Llama-2-7B-chat-hf} is the strongest evaluated model at 1k and 2k contexts (61.2\% and 56.8\%, respectively).}
\cl{On RepoBench-C, in contrast, diffusion LLMs are competitive or stronger at the evaluated 4k input length: \textsc{Dream-Coder-v0-Instruct-7B} reaches 32.2\% EM and 72.1\% ES, compared with 22.1\% and 62.8\% for \textsc{Llama-2-7B-chat-hf}.}
\cl{Thus, the shorter-input results do not uniformly favor either model family across all benchmarks and lengths.}

\cl{\textbf{At longer inputs, the evaluated AR baseline degrades more sharply than the three diffusion models.} On RepoQA, \textsc{Llama-2-7B-chat-hf} drops from 46.2\% at 3k to 9.0\% at 4k and 1.2\% at 8k. At 8k, the three diffusion models retain higher accuracy, although \textsc{LLaDA-1.5} and \textsc{Dream-Coder-v0-Instruct-7B} also decline materially to 14.8\% and 8.3\%. On RepoBench-C, the AR baseline drops from 22.1\% to 3.4\% EM and from 62.8\% to 19.2\% ES between 4k and 8k, whereas all three diffusion models maintain similar scores over this interval.}

\cl{\textbf{Robustness over increasing input length is not uniform across the evaluated diffusion LLMs.}}
\cl{At 8k, all three diffusion models retain nonzero performance, but the degree of degradation differs substantially by model and benchmark. Because their documented context capacities differ, this observation should not be interpreted as a uniform length-extrapolation test.}
\cl{On RepoQA, \textsc{DiffuCoder-7B-cpGRPO} is the most stable across the 1k--8k range, while \textsc{Dream-Coder-v0-Instruct-7B} retains less than 20\% of its short-input accuracy at 8k.}
On RepoBench-C, all three diffusion LLMs maintain or improve performance from 4k to 8k, but \textsc{Dream-Coder-v0-Instruct-7B} shows the strongest absolute scores throughout.
\cl{These results provide evidence of greater robustness over the evaluated input range for the three diffusion systems relative to this AR baseline. They should not be interpreted as a paradigm-wide property or a controlled effect of generation architecture.}

\begin{boxK}
\small \faIcon{pencil-alt} \textbf{Answer to \cl{RQ2}:}
\cc
\cl{On RepoQA and RepoBench-C, the three evaluated diffusion LLMs degrade more gracefully than one comparable-scale AR baseline over the common input-length range. Because the systems do not all have the same nominal context window, this finding is specific to the evaluated models and benchmarks rather than a fully context-controlled, paradigm-wide result.}
\end{boxK}

\end{cp}

\subsection{\cl{RQ3}: Impact of Diffusion LLM Settings on Effectiveness}
\label{sec:rq2}

\textbf{Motivation.}
The usability and quality of generated code are essential for real-world applications. In this section, we aim to investigate which factors influence the effectiveness of diffusion LLMs in code generation. We consider several factors that are expected to play an important role, including factors unique to diffusion LLMs such as the number of diffusion steps, the remasking strategy, the generation length, and whether to adopt block diffusion, as well as more general factors commonly studied in AR models such as decoding temperature.

\subsubsection{Impact of Generation Length} \label{sec:rq2_length}
\cl{Diffusion LLMs typically begin inference from a fully masked canvas of a predefined generation length, which sets the available output positions even though the semantic output may terminate earlier with \texttt{EOS}.} We believe this factor plays an important role in shaping the usability of the generated code.

\vspace{4pt} \noindent
\textbf{Setting.}
\begingroup\emergencystretch=2em
\cl{We evaluate five diffusion LLMs---\textsc{LLaDA-1.5}, \textsc{LLaDA-8B-Instruct}, \textsc{Dream-v0-Instruct-7B}, \textsc{Dream-Coder-v0-Instruct-7B}, and \textsc{DiffuCoder-7B-cpGRPO}---across HumanEval, MBPP, LiveCodeBench, and HumanEval-X. Where compatible results are available, we additionally include \textsc{DreamOn-v0-7B} and \textsc{Mercury-Coder-Small} on HumanEval and MBPP. We consider generation lengths from 64 to 2048, except on HumanEval-X, where the evaluated range is 64--1024. The exact model--length coverage is reported in Tables~\ref{tab:rq2_length} and~\ref{tab:rq2_length_new}; pass@1 is the evaluation metric.}
\par\endgroup

\vspace{4pt} \noindent
\textbf{Results and Analyses.}
\cl{The results are shown in Table~\ref{tab:rq2_length} for HumanEval, MBPP, and LiveCodeBench, and in Table~\ref{tab:rq2_length_new} for HumanEval-X.}

\begin{table}[ht]
\centering
\small
\caption{pass@1 (\%) across different generation lengths on HumanEval, MBPP, and LiveCodeBench.}
\label{tab:rq2_length}
\adjustbox{max width=\linewidth}{
\begin{tabular}{@{} l c c c c c c @{}}
\toprule
\textbf{Model} & \multicolumn{6}{c}{\textbf{Generation Length}} \\
\cmidrule(lr){2-7}
& 2048 & 1024 & 512 & 256 & 128 & 64 \\
\midrule
\multicolumn{7}{l}{\textbf{HumanEval}} \\
\textsc{LLaDA-8B-Instruct} & 45.7 & \textbf{47.0} & 45.7 & 42.7 & 39.6 & 30.5\\

\textsc{LLaDA-1.5} & 45.1 & 45.1 & 42.1 & \textbf{45.5} & 42.1 & 34.8\\

\textsc{Dream-v0-Instruct-7B} & 27.4 & 29.9 & \cl{41.5} & 32.3 & 34.8 & \textbf{43.9}\\

\textsc{Dream-Coder-v0-Instruct-7B} & 52.4 & 56.1 & 59.8 & 59.1 & \textbf{64.0} & 40.9\\

\textsc{DiffuCoder-7B-cpGRPO} & 61.0 & 61.0 & 61.6 & 61.6 & \textbf{65.2} & 54.9\\

\textsc{DreamOn-v0-7B} & -- & -- & 49.4 & 51.2 & \textbf{52.4} & 40.9\\

\textsc{Mercury-Coder-Small} & \textbf{86.0} & \textbf{86.0} & 84.8 & 65.9 & 2.4 & 2.4\\

\midrule
\multicolumn{7}{l}{\textbf{MBPP}} \\
\textsc{LLaDA-8B-Instruct} & 38.2 & 39.0 & 39.4 & \textbf{40.4} & 40.2 & 33.2\\

\textsc{LLaDA-1.5} & 38.4 & 38.4 & 40.6 & \textbf{42.0} & 40.0 & 31.6\\

\textsc{Dream-v0-Instruct-7B} & 42.0 & 46.0 & 45.2 & 43.4 & \textbf{46.8} & 39.0\\

\textsc{Dream-Coder-v0-Instruct-7B} & 53.8 & 57.6 & 59.2 & \textbf{59.6} & 40.0 & 41.6\\

\textsc{DiffuCoder-7B-cpGRPO} & 61.4 & 61.2 & 61.0 & 61.8 & \textbf{62.2} & 58.8\\

\textsc{DreamOn-v0-7B} & -- & -- & 53.2 & \textbf{54.0} & 50.8 & 44.8\\

\textsc{Mercury-Coder-Small} & 76.0 & \textbf{76.2} & 76.0 & 75.6 & 72.0 & 58.8\\

\midrule
\multicolumn{7}{l}{\textbf{LiveCodeBench}} \\
\textsc{LLaDA-8B-Instruct} & 5.6 & 4.7 & 6.2 & \textbf{7.9} & 3.8 & 3.8\\

\textsc{LLaDA-1.5} & 6.2 & 5.3 & 6.2 & \textbf{7.3} & 5.0 & 3.2\\

\textsc{Dream-v0-Instruct-7B} & 1.2 & 1.2 & 1.8 & 2.4 & \textbf{3.8} & 3.2\\

\textsc{Dream-Coder-v0-Instruct-7B} & 5.3 & 5.3 & 5.3 & 6.7 & \textbf{7.3} & 4.1\\

\textsc{DiffuCoder-7B-cpGRPO} & 6.2 & \textbf{6.5} & \textbf{6.5} & \textbf{6.5} & \textbf{6.5} & 5.6\\

\bottomrule
\end{tabular}
}
\end{table}

\begin{table}[ht]
\centering
\small
\caption{pass@1 (\%) across different generation lengths on HumanEval-X.}
\label{tab:rq2_length_new}
\adjustbox{max width=\linewidth}{
\begin{tabular}{@{} l c c c c c @{}}
\toprule
\textbf{Model} & \multicolumn{5}{c}{\textbf{Generation Length}} \\
\cmidrule(lr){2-6}
& 1024 & 512 & 256 & 128 & 64 \\
\midrule
\multicolumn{6}{l}{\textbf{HumanEval-X (C++)}} \\
\textsc{LLaDA-8B-Instruct} & 28.7 & 25.6 & \textbf{29.3} & 25.0 & 20.1\\

\textsc{LLaDA-1.5} & 28.0 & 28.7 & \textbf{29.9} & 27.4 & 18.9\\

\textsc{Dream-v0-Instruct-7B} & 11.6 & 15.2 & 14.6 & \textbf{23.8} & 23.2\\

\textsc{Dream-Coder-v0-Instruct-7B} & 28.0 & 30.5 & \textbf{51.2} & 50.0 & 23.8\\

\textsc{DiffuCoder-7B-cpGRPO} & 32.9 & 32.9 & 33.5 & \textbf{34.1} & 33.5\\

\midrule
\multicolumn{6}{l}{\textbf{HumanEval-X (Java)}} \\
\textsc{LLaDA-8B-Instruct} & 34.8 & \textbf{42.1} & 33.5 & 31.7 & 17.1\\

\textsc{LLaDA-1.5} & 37.2 & 36.6 & \textbf{37.8} & 29.3 & 15.2\\

\textsc{Dream-v0-Instruct-7B} & 28.7 & 27.4 & 28.0 & \textbf{40.2} & 32.9\\

\textsc{Dream-Coder-v0-Instruct-7B} & 39.0 & 41.5 & 53.7 & \textbf{54.3} & 24.4\\

\textsc{DiffuCoder-7B-cpGRPO} & \textbf{40.2} & 39.6 & 37.2 & 38.4 & 32.9\\

\midrule
\multicolumn{6}{l}{\textbf{HumanEval-X (Go)}} \\
\textsc{LLaDA-8B-Instruct} & \textbf{31.7} & 28.0 & 22.6 & 22.6 & 12.2\\

\textsc{LLaDA-1.5} & 25.0 & \textbf{26.8} & 20.7 & 23.2 & 11.0\\

\textsc{Dream-v0-Instruct-7B} & 9.8 & 13.4 & 11.6 & \textbf{20.1} & 18.3\\

\textsc{Dream-Coder-v0-Instruct-7B} & 28.7 & 29.9 & 36.6 & \textbf{45.1} & 21.3\\

\textsc{DiffuCoder-7B-cpGRPO} & 25.6 & 25.6 & 23.8 & 25.6 & \textbf{26.2}\\

\midrule
\multicolumn{6}{l}{\textbf{HumanEval-X (JavaScript)}} \\
\textsc{LLaDA-8B-Instruct} & 20.1 & \textbf{25.6} & 22.0 & 20.1 & 21.3\\

\textsc{LLaDA-1.5} & 28.0 & \textbf{32.9} & 25.6 & 29.9 & 21.3\\

\textsc{Dream-v0-Instruct-7B} & 21.3 & 23.2 & 23.8 & \textbf{35.4} & 32.3\\

\textsc{Dream-Coder-v0-Instruct-7B} & 41.5 & 41.5 & \textbf{51.8} & 50.0 & 30.5\\

\textsc{DiffuCoder-7B-cpGRPO} & 45.1 & 47.6 & 47.0 & \textbf{48.2} & 42.1\\

\bottomrule
\end{tabular}
}
\end{table}

\begin{figure}[!htbp]
    \centering
    \includegraphics[width=0.55\linewidth]{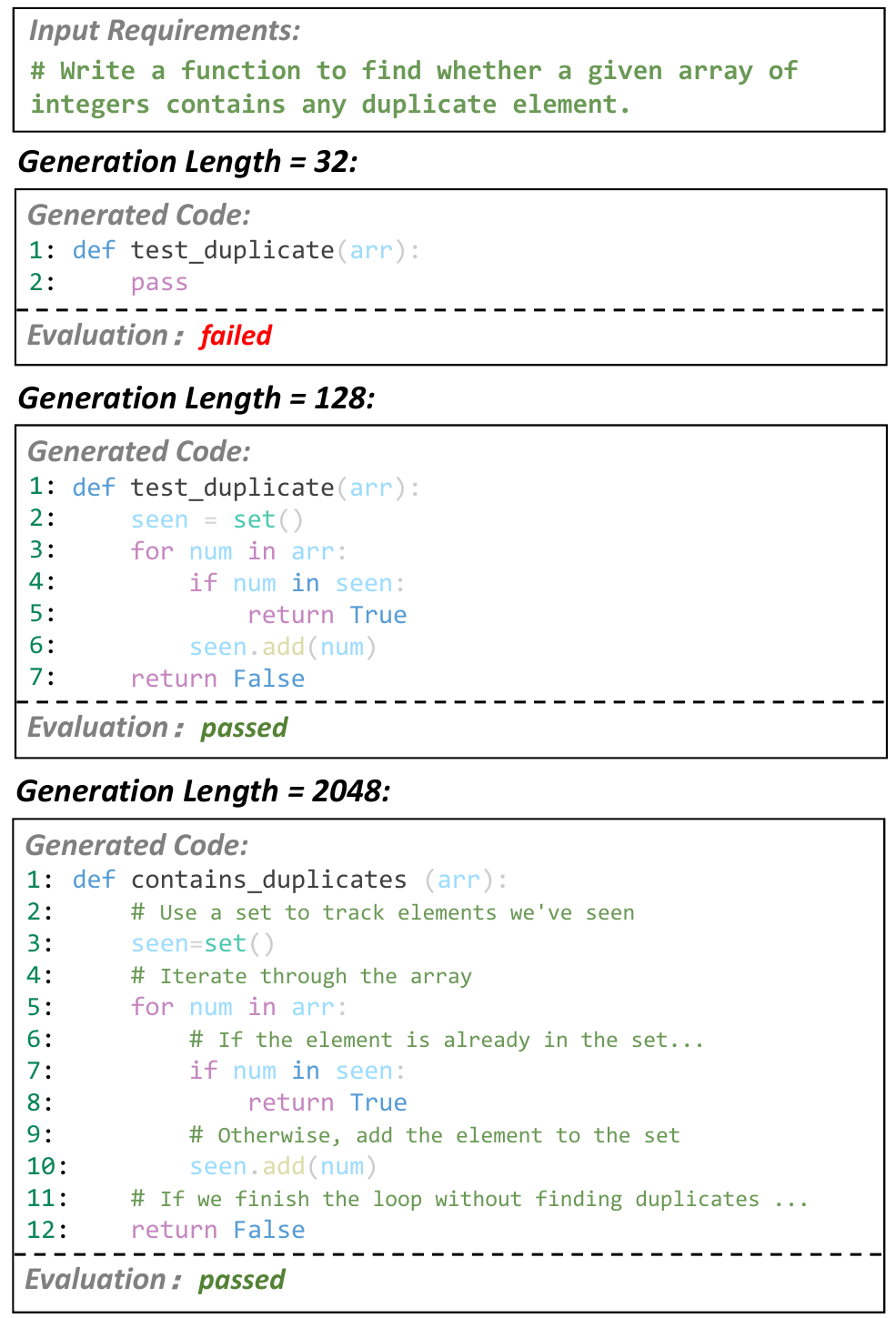}
    \caption{Case study illustrating the impact of different generation lengths on code usability in diffusion LLMs.}
    \label{fig:case_length}
    \Description{\cl{A code-generation example at lengths 32, 128, and 2048, showing an incomplete short output, a passing intermediate-length output, and a longer output with additional commentary.}}
\end{figure}

\cl{\textbf{Generation length often has a non-monotonic relationship with effectiveness, although the shape varies by model and benchmark.}}
For instance, on HumanEval, the pass@1 score of \textsc{Dream-Coder-v0-Instruct-7B} reaches its peak \cl{(64.0\%)} when the generation length expands from 64 \cl{(40.9\%)} to 128, \cl{and then generally declines to 52.4\% by length 2048, with a small increase at length 512}. \cl{Short sequences can lack sufficient capacity to produce complete code, while longer sequences provide room for additional tokens.}
\cl{Prior work has suggested that excessively long generation lengths can reduce output quality~\cite{longllada}. Figure~\ref{fig:case_length} illustrates one evaluated case in which the extra content consists largely of comments rather than faulty code; this example does not establish a general mechanism.}

\textbf{The optimal generation length varies across models.} For example, on HumanEval, the best setting for \textsc{LLaDA-8B-Instruct} is 1024, whereas for \textsc{DiffuCoder-7B-cpGRPO} the best length is only 128. This shows that diffusion LLMs differ considerably in their optimal generation length.

\cl{\textbf{The best generation length can also vary across benchmarks.} For example, \textsc{Dream-Coder-v0-Instruct-7B} shifts from a best length of 128 on HumanEval to 256 on MBPP, whereas \textsc{DiffuCoder-7B-cpGRPO} peaks at 128 on both benchmarks. Differences in task format and expected output length may contribute to this benchmark dependence, but the experiments do not isolate their individual effects.}

\cl{On LiveCodeBench, \textsc{LLaDA-8B-Instruct} achieves its best result (7.9\%) at a length of 256, while \textsc{DiffuCoder-7B-cpGRPO} varies within 5.6\%--6.5\% across the evaluated lengths. On HumanEval-X, the best length likewise varies by model and language. For example, \textsc{Dream-Coder-v0-Instruct-7B} peaks at 256 on C++ (51.2\%), at 128 on Java (54.3\%), at 128 on Go (45.1\%), and at 256 on JavaScript (51.8\%). Thus, intermediate lengths are often effective, but the optimum is configuration-specific rather than universal.}

\begin{figure}[t]
    \centering
    \includegraphics[width=\linewidth]{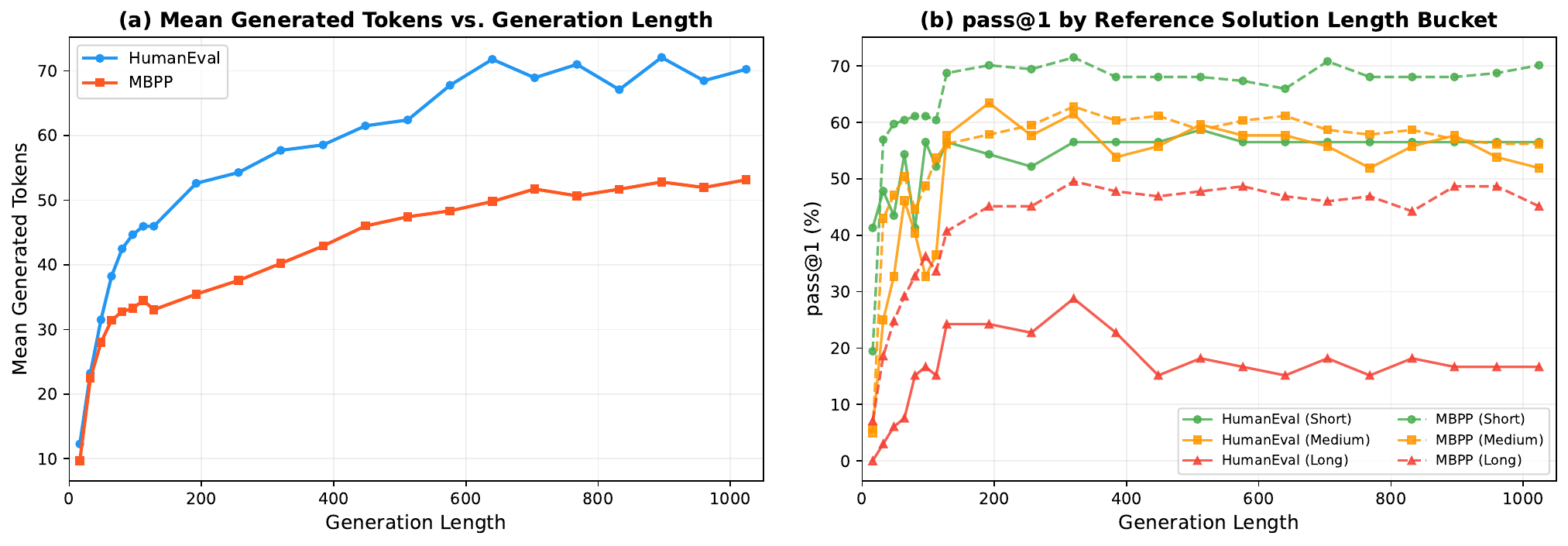}
    \caption{Left: Mean number of generated tokens as a function of the generation length on HumanEval and MBPP. Right: pass@1 (\%) stratified by the reference solution length (Short $\leq$ 26 tokens, Medium 27--45 tokens, Long $>$ 45 tokens), evaluated under the same generation length sweep.}
    \label{fig:genlen_bucket}
    \Description{\cl{Generated-token counts and pass-at-one trends across generation lengths and reference-solution length buckets.}}
\end{figure}

\cl{\textbf{Disentangling the effect of task output length.} To better understand whether the observed trend is driven by the mixture of tasks with different required output lengths, we conduct a fine-grained ablation on a representative diffusion LLM, \textsc{LLaDA-8B-Instruct}.}
\cl{We sweep the generation length from 16 to 1024 at a granularity of 16--64 tokens (22 settings in total) on HumanEval and MBPP.}
\cl{For each generation length, we record the mean number of generated tokens (Figure~\ref{fig:genlen_bucket}, left) and the pass@1 performance.}
\cl{We then bucket the tasks by the token count of their reference solution into three groups: Short ($\leq$26 tokens), Medium (27--45 tokens), and Long ($>$45 tokens), and examine the pass@1 trend within each bucket (Figure~\ref{fig:genlen_bucket}, right).}

\cl{Two observations emerge. First, the mean generated-token count plateaus once the generation canvas exceeds the typical output length, which is consistent with trailing \texttt{EOS} positions contributing to the efficiency penalty at large lengths.}
\cl{Second, reference-solution length interacts with the preferred generation length, although the exact optimum remains benchmark-dependent. Short tasks peak at length 512 on HumanEval and 320 on MBPP, while the Long-task buckets peak at 320 and 512, respectively.}
\cl{For instance, on HumanEval, Long tasks reach only 7.6\% pass@1 at length 64 and improve to 24.2\% at length 128, peaking at 28.8\% at length 320, whereas Short tasks already achieve 54.3\% at length 64 and peak at 58.7\% at length 512.}
\cl{On MBPP, the same pattern holds: Long tasks improve from 29.2\% at length 64 to 47.8\% at length 512, while Short tasks achieve 60.4\% at length 64 and peak at 71.5\% at length 320.}
\cl{The within-bucket curves retain some non-monotonicity, so task-length heterogeneity does not fully explain the aggregate pattern. However, their different optima also show that task composition affects the observed trend.}
\cl{The practical implication is that generation length should be selected jointly for the model and expected task distribution, rather than treated as a universally optimal constant.}

\subsubsection{Impact of Diffusion Steps}
\label{sec:rq2_step}
\cl{A distinctive feature of diffusion LLMs is the ability to control the number of denoising iterations. For a fixed generation canvas, fewer steps commit more token predictions per iteration on average, whereas more steps permit finer-grained iterative updates.}

\vspace{4pt} \noindent
\textbf{Setting.}
\cl{We evaluate the five diffusion LLMs shared across all four benchmarks in the generation-length analysis, using 8, 16, 32, 64, 128, 256, and 512 diffusion steps. Pass@1 is the metric, and the generation length is fixed at 512.}

\vspace{4pt} \noindent
\textbf{Results and Analyses.}
\cl{The results are shown in Table~\ref{tab:rq2_step} for HumanEval and MBPP and in Table~\ref{tab:rq2_step_new} for LiveCodeBench and HumanEval-X.}

\begin{table}[ht]
\centering
\small
\caption{pass@1 (\%) across different diffusion steps on HumanEval and MBPP.}
\label{tab:rq2_step}
\adjustbox{max width=\linewidth}{
\begin{tabular}{@{} l c c c c c c c @{}}
\toprule
\textbf{Model} & \multicolumn{7}{c}{\textbf{Diffusion Steps}} \\
\cmidrule(lr){2-8}
& 512 & 256 & 128 & 64 & 32 & 16 & 8 \\
\midrule
\multicolumn{8}{l}{\textbf{HumanEval}} \\
\textsc{LLaDA-8B-Instruct} & \textbf{45.7} & 34.1 & 22.6 & 7.9 & 3.7 & 3.7 & 0.6\\

\textsc{LLaDA-1.5} & \textbf{42.1} & 34.8 & 18.9 & 9.1 & 4.9 & 1.8 & 1.2\\

\textsc{Dream-v0-Instruct-7B} & \textbf{41.5} & 34.8 & 34.8 & 19.5 & 6.7 & 1.8 & 1.2\\

\textsc{Dream-Coder-v0-Instruct-7B} & \textbf{59.8} & 57.9 & 55.5 & 40.9 & 23.2 & 14.6 & 12.2\\

\textsc{DiffuCoder-7B-cpGRPO} & 61.6 & \textbf{62.2} & 51.8 & 41.5 & 38.4 & 33.5 & 28.7\\

\midrule
\multicolumn{8}{l}{\textbf{MBPP}} \\
\textsc{LLaDA-8B-Instruct} & \textbf{39.4} & 31.2 & 21.6 & 11.2 & 3.8 & 0.0 & 0.0\\

\textsc{LLaDA-1.5} & \textbf{40.6} & 33.0 & 20.4 & 10.4 & 3.2 & 0.2 & 0.2\\

\textsc{Dream-v0-Instruct-7B} & \textbf{45.2} & 43.8 & 32.2 & 22.4 & 8.4 & 5.0 & 0.6\\

\textsc{Dream-Coder-v0-Instruct-7B} & \textbf{59.2} & 56.2 & 52.6 & 43.4 & 29.2 & 17.6 & 12.2\\

\textsc{DiffuCoder-7B-cpGRPO} & \textbf{61.0} & 60.0 & 53.8 & 43.8 & 33.4 & 29.2 & 25.4\\

\bottomrule
\end{tabular}
}
\end{table}

\begin{table}[ht]
\centering
\small
\caption{pass@1 (\%) across different diffusion steps on LiveCodeBench and HumanEval-X.}
\label{tab:rq2_step_new}
\adjustbox{max width=\linewidth}{
\begin{tabular}{@{} l c c c c c c c @{}}
\toprule
\textbf{Model} & \multicolumn{7}{c}{\textbf{Diffusion Steps}} \\
\cmidrule(lr){2-8}
& 512 & 256 & 128 & 64 & 32 & 16 & 8 \\
\midrule
\multicolumn{8}{l}{\textbf{LiveCodeBench}} \\
\textsc{LLaDA-8B-Instruct} & \textbf{6.2} & 5.3 & 0.6 & 0.0 & 0.0 & 0.0 & 0.0\\

\textsc{LLaDA-1.5} & \textbf{6.2} & 5.3 & 0.9 & 0.3 & 0.0 & 0.0 & 0.0\\

\textsc{Dream-v0-Instruct-7B} & 1.8 & \textbf{2.9} & 2.1 & 1.5 & 0.3 & 0.0 & 0.0\\

\textsc{Dream-Coder-v0-Instruct-7B} & 5.3 & \textbf{6.5} & 5.6 & 2.4 & 2.0 & 1.2 & 0.0\\

\textsc{DiffuCoder-7B-cpGRPO} & \textbf{6.5} & 5.6 & 5.6 & 4.7 & 3.2 & 3.2 & 2.0\\

\midrule
\multicolumn{8}{l}{\textbf{HumanEval-X (C++)}} \\
\textsc{LLaDA-8B-Instruct} & \textbf{25.6} & 19.5 & 9.8 & 1.8 & 1.2 & 0.0 & 0.0\\

\textsc{LLaDA-1.5} & \textbf{28.7} & 25.6 & 16.5 & 4.3 & 1.2 & 0.0 & 0.0\\

\textsc{Dream-v0-Instruct-7B} & 15.9 & \textbf{21.3} & 15.9 & 8.5 & 2.4 & 1.8 & 1.2\\

\textsc{Dream-Coder-v0-Instruct-7B} & 30.5 & \textbf{31.1} & 28.7 & 26.8 & 12.8 & 4.3 & 1.8\\

\textsc{DiffuCoder-7B-cpGRPO} & 32.9 & \textbf{34.1} & 31.7 & 19.5 & 12.2 & 5.5 & 4.3\\

\midrule
\multicolumn{8}{l}{\textbf{HumanEval-X (Java)}} \\
\textsc{LLaDA-8B-Instruct} & \textbf{42.1} & 28.0 & 16.5 & 10.4 & 1.8 & 0.6 & 0.6\\

\textsc{LLaDA-1.5} & \textbf{36.6} & 29.3 & 20.7 & 4.9 & 0.0 & 0.0 & 0.0\\

\textsc{Dream-v0-Instruct-7B} & 27.4 & \textbf{32.3} & 25.6 & 20.7 & 13.4 & 5.5 & 3.7\\

\textsc{Dream-Coder-v0-Instruct-7B} & 41.5 & \textbf{42.7} & 36.0 & 31.7 & 18.3 & 9.8 & 7.3\\

\textsc{DiffuCoder-7B-cpGRPO} & \textbf{39.6} & 39.0 & 32.9 & 25.6 & 18.3 & 17.1 & 13.4\\

\midrule
\multicolumn{8}{l}{\textbf{HumanEval-X (Go)}} \\
\textsc{LLaDA-8B-Instruct} & \textbf{28.0} & 20.7 & 12.8 & 5.5 & 1.8 & 1.8 & 1.8\\

\textsc{LLaDA-1.5} & \textbf{26.8} & 18.9 & 12.2 & 5.5 & 1.2 & 1.2 & 1.2\\

\textsc{Dream-v0-Instruct-7B} & 13.4 & \textbf{15.2} & \textbf{15.2} & 8.5 & 2.4 & 0.6 & 0.6\\

\textsc{Dream-Coder-v0-Instruct-7B} & 29.9 & \textbf{32.3} & 29.3 & 20.1 & 15.2 & 11.0 & 4.9\\

\textsc{DiffuCoder-7B-cpGRPO} & 25.6 & \textbf{26.8} & 22.0 & 15.9 & 10.4 & 6.1 & 4.3\\

\midrule
\multicolumn{8}{l}{\textbf{HumanEval-X (JavaScript)}} \\
\textsc{LLaDA-8B-Instruct} & \textbf{25.6} & 20.7 & 7.3 & 5.5 & 1.2 & 0.6 & 0.6\\

\textsc{LLaDA-1.5} & \textbf{32.9} & 22.6 & 13.4 & 7.3 & 2.4 & 0.6 & 0.6\\

\textsc{Dream-v0-Instruct-7B} & 23.2 & \textbf{28.0} & 25.0 & 18.3 & 9.8 & 4.9 & 4.3\\

\textsc{Dream-Coder-v0-Instruct-7B} & \textbf{41.5} & 40.2 & 40.2 & 31.1 & 21.3 & 10.4 & 7.3\\

\textsc{DiffuCoder-7B-cpGRPO} & \textbf{47.6} & 45.7 & 37.8 & 30.5 & 24.4 & 19.5 & 14.6\\

\bottomrule
\end{tabular}
}
\end{table}

\textbf{The usability of generated code generally improves as the number of diffusion steps increases.}
\cl{At 8 steps, every evaluated model performs below its high-step setting, although the magnitude of the drop varies considerably. The best result usually occurs at 256 or 512 steps, depending on the model and benchmark.}

\begingroup\emergencystretch=2em
\cl{\textbf{Sensitivity at high step counts remains model-dependent.} On HumanEval, increasing the number of steps from 256 to 512 changes pass@1 by only 1.9 points for \textsc{Dream-Coder-v0-Instruct-7B} and by $-0.6$ points for \textsc{DiffuCoder-7B-cpGRPO}, but improves \textsc{LLaDA-8B-Instruct} and \textsc{LLaDA-1.5} by 11.6 and 7.3 points, respectively. The reason for this model-dependent sensitivity remains unclear.}
\par\endgroup

\cl{On LiveCodeBench, high step counts are particularly important for some models. For instance, \textsc{LLaDA-8B-Instruct} drops from 6.2\% at 512 steps to 0.6\% at 128 steps and 0.0\% at 64 steps. Across HumanEval-X, pass@1 likewise tends to increase toward higher step counts, although several Dream and DiffuCoder configurations peak at 256 steps. For example, \textsc{Dream-Coder-v0-Instruct-7B} improves from 1.8\% at 8 steps to 30.5\% at 512 steps on C++, and from 7.3\% to 41.5\% on Java.}

\subsubsection{Impact of Remasking Strategy} \label{sec:rq2_remask}
During each inference step, every \texttt{[MASK]} token produces a predicted token. However, only a subset of these predictions is retained, while the remaining positions are remasked for further refinement.

\vspace{4pt} \noindent
\textbf{Setting.}
We evaluate two widely used remasking strategies: random remasking and low-confidence remasking. Random remasking assigns mask positions uniformly at random in each step, whereas low-confidence remasking selectively masks tokens with lower prediction confidence.
\cl{We evaluate the same five models and four benchmarks as in the diffusion-step analysis, with pass@1 as the metric and generation length fixed at 512.}

\vspace{4pt} \noindent
\textbf{Results and Analyses.}
\cl{The results on HumanEval, MBPP, and LiveCodeBench are shown in Table~\ref{tab:rq2_remask}, while additional results on HumanEval-X are provided in Table~\ref{tab:rq2_remask_new}.}

\begin{table}[ht]
\centering
\small
\caption{pass@1 (\%) across different remasking strategies on HumanEval, MBPP, and LiveCodeBench.}
\label{tab:rq2_remask}
\setlength{\tabcolsep}{4pt}
\adjustbox{max width=\linewidth}{
\begin{tabular}{@{} l c c c c @{}}
\toprule
\textbf{Model} & \textbf{Strategy} & \textbf{HumanEval} & \textbf{MBPP} & \textbf{LiveCodeBench} \\
\midrule
\textsc{LLaDA-8B-Instruct} & Random Selection & 11.6 & 17.4 & 1.8 \\
& Confidence-Based & \textbf{43.9} \textcolor{red}{($\uparrow$32.3)} & \textbf{38.2} \textcolor{red}{($\uparrow$20.8)} & \textbf{7.6} \textcolor{red}{($\uparrow$5.8)} \\
\textsc{LLaDA-1.5} & Random Selection & 15.2 & 16.8 & 0.6 \\
& Confidence-Based & \textbf{48.2} \textcolor{red}{($\uparrow$33.0)} & \textbf{40.6} \textcolor{red}{($\uparrow$23.8)} & \textbf{7.0} \textcolor{red}{($\uparrow$6.4)} \\
\textsc{Dream-v0-Instruct-7B} & Random Selection & 20.7 & 28.2 & 1.2 \\
& Confidence-Based & \textbf{43.3} \textcolor{red}{($\uparrow$22.6)} & \textbf{51.6} \textcolor{red}{($\uparrow$23.4)} & \textbf{5.3} \textcolor{red}{($\uparrow$4.1)} \\
\textsc{Dream-Coder-v0-Instruct-7B} & Random Selection & 34.1 & 31.0 & 1.8 \\
& Confidence-Based & \textbf{69.5} \textcolor{red}{($\uparrow$35.4)} & \textbf{59.2} \textcolor{red}{($\uparrow$28.2)} & \textbf{10.0} \textcolor{red}{($\uparrow$8.2)} \\
\textsc{DiffuCoder-7B-cpGRPO} & Random Selection & 40.2 & 46.6 & 2.9 \\
& Confidence-Based & \textbf{69.5} \textcolor{red}{($\uparrow$29.3)} & \textbf{61.0} \textcolor{red}{($\uparrow$14.4)} & \textbf{7.3} \textcolor{red}{($\uparrow$4.4)} \\
\bottomrule
\end{tabular}
}
\end{table}

\begin{figure}[t]
    \centering
    \includegraphics[width=0.34\linewidth]{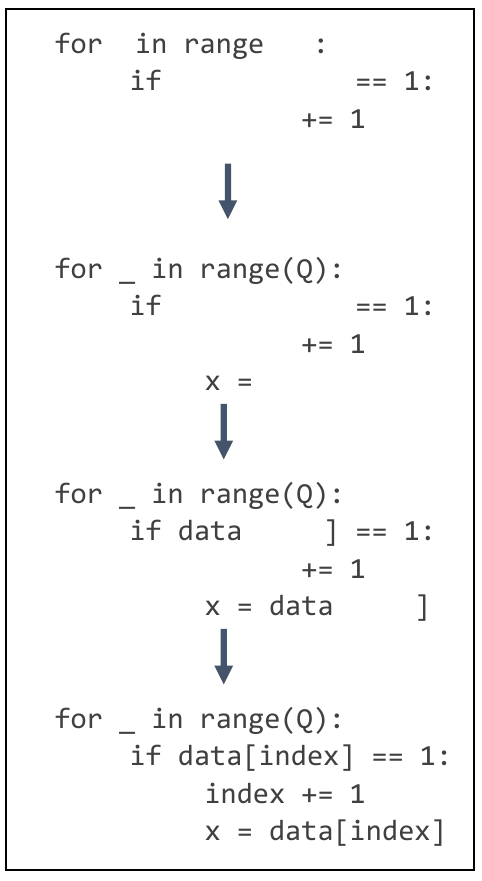}
    \caption{\cl{Case study of \textsc{Dream-Coder-v0-Instruct-7B} on a LiveCodeBench problem, illustrating the progressive refinement behavior of low-confidence remasking.}}
    \label{fig:case_remask}
    \Description{\cl{Four intermediate decoding states in which masked code positions are progressively filled under low-confidence remasking.}}
\end{figure}

\cl{\textbf{Confidence-based remasking consistently outperforms random remasking in the evaluated configurations.}}
\cl{For example, on HumanEval, \textsc{LLaDA-1.5} and \textsc{Dream-Coder-v0-Instruct-7B} achieve improvements of 33.0 and 35.4 percentage points, respectively, when using low-confidence remasking instead of random remasking. On LiveCodeBench, \textsc{LLaDA-8B-Instruct} improves from 1.8\% (random) to 7.6\% (confidence-based), and \textsc{Dream-Coder-v0-Instruct-7B} improves from 1.8\% to 10.0\%. Across HumanEval-X, the benefits are also consistent: \textsc{Dream-Coder-v0-Instruct-7B} improves from 16.5\% to 54.3\% on C++ and from 24.4\% to 63.4\% on Java. In the representative case in Figure~\ref{fig:case_remask}, low-confidence remasking first forms a coarse code structure and then refines implementation details. The sequence develops from placeholder variables and a partial loop into a syntactically complete solution. This example is consistent with coarse-to-fine refinement, but is illustrative rather than evidence that the mechanism holds generally.}

\cl{\textbf{Statistical significance.} To validate that the improvements of confidence-based remasking over random remasking are not due to chance, we conduct McNemar tests on the per-problem pass/fail outcomes for representative models across benchmarks.}
\cl{For \textsc{LLaDA-8B-Instruct} and \textsc{Dream-v0-Instruct-7B}, the confidence-based strategies consistently yield statistically significant improvements over the baselines on HumanEval, MBPP, LiveCodeBench, and all four languages of HumanEval-X (all \mbox{$p < 0.01$}).}
\cl{For instance, comparing low-confidence to random remasking for \textsc{LLaDA-8B-Instruct} on HumanEval yields $p < 0.001$ ($a_{10}=57$ problems improved, $a_{01}=4$ degraded), and on MBPP $p < 0.001$ ($a_{10}=123$, $a_{01}=19$).}
\cl{Full McNemar test results are provided in the supplementary material.}

\begin{table}[ht]
\centering
\small
\caption{pass@1 (\%) across different remasking strategies on HumanEval-X.}
\label{tab:rq2_remask_new}
\adjustbox{max width=\linewidth}{
\begin{tabular}{@{} l c c c c c @{}}
\toprule
\textbf{Model} & \textbf{Strategy} & \textbf{C++} & \textbf{Java} & \textbf{Go} & \textbf{JavaScript} \\
\midrule
\textsc{LLaDA-8B-Instruct} & Random Selection & 8.5 & 10.4 & 3.7 & 7.9 \\
& Confidence-Based & \textbf{26.8} & \textbf{41.5} & \textbf{30.5} & \textbf{26.2} \\
\textsc{LLaDA-1.5} & Random Selection & 9.8 & 11.6 & 6.1 & 6.1 \\
& Confidence-Based & \textbf{29.9} & \textbf{34.8} & \textbf{24.4} & \textbf{33.5} \\
\textsc{Dream-v0-Instruct-7B} & Random Selection & 14.6 & 22.0 & 8.5 & 23.2 \\
& Confidence-Based & \textbf{26.2} & \textbf{40.9} & \textbf{28.0} & \textbf{49.4} \\
\textsc{Dream-Coder-v0-Instruct-7B} & Random Selection & 16.5 & 24.4 & 9.1 & 25.6 \\
& Confidence-Based & \textbf{54.3} & \textbf{63.4} & \textbf{53.7} & \textbf{54.9} \\
\textsc{DiffuCoder-7B-cpGRPO} & Random Selection & 34.1 & 28.7 & 16.5 & 39.0 \\
& Confidence-Based & \textbf{43.3} & \textbf{51.2} & \textbf{42.1} & \textbf{55.5} \\
\bottomrule
\end{tabular}
}
\end{table}

\FloatBarrier

\subsubsection{Impact of Block Length} \label{sec:rq2_block}
\cl{Some diffusion LLMs such as \textsc{LLaDA-8B-Instruct} support block diffusion, which divides the sequence into blocks decoded sequentially from left to right while applying diffusion decoding within each block.}

\vspace{4pt} \noindent
\textbf{Setting.}
We evaluate \textsc{LLaDA-8B-Instruct} and \textsc{LLaDA-1.5} with generation length fixed at 512, varying block length from 1 to 512. \cl{Experiments are conducted on HumanEval, MBPP, LiveCodeBench, and HumanEval-X} using pass@1 as the metric.

\vspace{4pt} \noindent
\textbf{Results and Analyses.}
\cl{The complete results are shown in Table~\ref{tab:rq2_block} for all benchmarks.}

\begin{table}[ht]
\centering
\small
\caption{pass@1 (\%) across different block lengths on HumanEval, MBPP, LiveCodeBench, and HumanEval-X.}
\label{tab:rq2_block}
\adjustbox{max width=\linewidth}{
\begin{tabular}{@{} l c c c c c c c c c c @{}}
\toprule
\textbf{Model} & \multicolumn{10}{c}{\textbf{Block Length}} \\
\cmidrule(lr){2-11}
& 1 & 2 & 4 & 8 & 16 & 32 & 64 & 128 & 256 & 512 \\
\midrule
\multicolumn{11}{l}{\textbf{HumanEval}} \\
\textsc{LLaDA-1.5} & 40.2 & 41.5 & \textbf{45.7} & 43.3 & 43.9 & 43.9 & 42.1 & 42.7 & 32.3 & 21.3\\

\textsc{LLaDA-8B-Instruct} & 40.9 & 42.1 & 42.1 & 44.5 & \textbf{46.3} & 43.9 & 43.3 & 38.4 & 20.1 & 17.1\\

\midrule
\multicolumn{11}{l}{\textbf{MBPP}} \\
\textsc{LLaDA-1.5} & 40.0 & \textbf{41.0} & 38.2 & 39.2 & 39.2 & 39.8 & 40.6 & 40.6 & 39.0 & 35.2\\

\textsc{LLaDA-8B-Instruct} & 38.8 & 39.4 & 38.4 & 38.0 & 39.2 & 39.4 & 39.6 & \textbf{39.8} & 33.2 & 28.0\\

\midrule
\multicolumn{11}{l}{\textbf{LiveCodeBench}} \\
\textsc{LLaDA-8B-Instruct} & \textbf{7.9} & 6.7 & 5.9 & 6.5 & 6.2 & 5.6 & 6.2 & 5.6 & 3.2 & 1.5\\

\textsc{LLaDA-1.5} & \textbf{8.2} & 7.9 & 7.0 & 7.6 & 7.6 & 6.7 & 6.2 & 6.7 & 5.9 & 3.5\\

\midrule
\multicolumn{11}{l}{\textbf{HumanEval-X (C++)}} \\
\textsc{LLaDA-8B-Instruct} & 26.8 & 26.8 & 26.8 & 26.2 & \textbf{27.4} & \textbf{27.4} & 25.6 & 25.0 & 19.5 & 5.5\\

\textsc{LLaDA-1.5} & 25.6 & 28.0 & 28.0 & 25.6 & 27.4 & 26.8 & \textbf{28.7} & 28.0 & 26.8 & 18.9\\

\midrule
\multicolumn{11}{l}{\textbf{HumanEval-X (Java)}} \\
\textsc{LLaDA-8B-Instruct} & 38.4 & 39.0 & 38.4 & 39.6 & 39.6 & 39.0 & \textbf{42.1} & 40.9 & 39.6 & 21.3\\

\textsc{LLaDA-1.5} & 37.8 & \textbf{38.4} & \textbf{38.4} & 33.5 & 37.2 & 37.8 & 36.6 & \textbf{38.4} & \textbf{38.4} & 36.0\\

\midrule
\multicolumn{11}{l}{\textbf{HumanEval-X (Go)}} \\
\textsc{LLaDA-8B-Instruct} & 23.8 & 24.4 & 24.4 & 27.4 & 23.8 & 25.6 & \textbf{28.0} & \textbf{28.0} & 18.3 & 8.5\\

\textsc{LLaDA-1.5} & 26.8 & 26.8 & 27.4 & \textbf{28.7} & 28.0 & 26.2 & 26.8 & 26.2 & 23.8 & 14.6\\

\midrule
\multicolumn{11}{l}{\textbf{HumanEval-X (JavaScript)}} \\
\textsc{LLaDA-8B-Instruct} & 24.4 & 23.2 & 23.2 & \textbf{26.8} & \textbf{26.8} & 24.4 & 25.6 & 25.6 & 22.0 & 12.2\\

\textsc{LLaDA-1.5} & 29.3 & 28.0 & 28.7 & 30.5 & 32.9 & 33.5 & 32.9 & 33.5 & \textbf{34.8} & 26.8\\

\bottomrule
\end{tabular}
}
\end{table}

\cl{\textbf{Very long block lengths generally reduce effectiveness, whereas the effect of short and intermediate blocks is benchmark-dependent.} For example, on HumanEval, \textsc{LLaDA-8B-Instruct} peaks at 46.3\% with a block length of 16, compared with 40.9\% at block length 1 and 17.1\% at block length 512. On MBPP, the curve is flatter: \textsc{LLaDA-8B-Instruct} achieves 38.8\% at block length 1, 39.2\% at block length 16, and 28.0\% at block length 512.}
\cl{The contrast between HumanEval and LiveCodeBench shows that no single intermediate block length is uniformly optimal; the present results do not isolate the mechanism behind this benchmark dependence.}

\cl{On LiveCodeBench, both models perform best at block length 1 and generally decline as block length increases; \textsc{LLaDA-8B-Instruct}, for example, drops from 7.9\% at block length 1 to 1.5\% at 512. Across HumanEval-X, the best block length varies by model and language, while the largest block length usually performs worse than the best intermediate setting.}

\subsubsection{Impact of Decoding Temperature} \label{sec:rq2_temperature}
Temperature is a parameter shared by both AR and diffusion LLMs that controls the level of randomness in the generated outputs. Higher temperatures generally produce more diverse results.

\vspace{4pt} \noindent
\textbf{Setting.}
\cl{We evaluate the five diffusion LLMs shared across HumanEval, LiveCodeBench, and HumanEval-X, and additionally report \textsc{DreamOn-v0-7B} on HumanEval. For a HumanEval reference comparison reported in the text, we inspect two AR baselines: \textsc{Seed-Coder-8B-Instruct} and \textsc{CodeLlama-7B-Instruct-hf}.}
\cl{The temperature is gradually increased from 0 to 2, and the diffusion models are evaluated on HumanEval, LiveCodeBench, and HumanEval-X using pass@5 and average pass@1 as metrics.}
For both pass@5 and pass@1, we repeat each configuration five times and report the mean and standard deviation.

\vspace{4pt} \noindent
\textbf{Results and Analyses.}
\cl{The diffusion-model results on HumanEval and LiveCodeBench are shown in Table~\ref{tab:rq2_temperature}. The results on HumanEval-X are presented in Table~\ref{tab:rq2_temperature_new}.}

\begin{table}[!t]
\centering
\small
\caption{\cl{pass@5 (\%) and pass@1 (\%) across different temperatures on HumanEval and LiveCodeBench. Parenthesized values report standard deviations over five runs; deterministic entries may omit a zero standard deviation.}}
\label{tab:rq2_temperature}

\adjustbox{max width=\linewidth}{
\begin{tabular}{@{} l l c c c c c c c @{}}
\toprule
\textbf{Model} & \textbf{Metric} & \multicolumn{7}{c}{\textbf{Temperature}} \\
\cmidrule(lr){3-9}
& & 0.0 & 0.2 & 0.4 & 0.8 & 1.2 & 1.6 & 2.0 \\
\midrule
\multicolumn{9}{l}{\textbf{HumanEval}} \\
\textsc{LLaDA-8B-Instruct} & pass@5 & 45.7 & 48.8(0.7) & 52.6(1.1) & 56.6(2.3) & \textbf{59.6(1.7)} & 49.8(3.6) & 13.7(1.1) \\
& pass@1 & \textbf{45.7} & 43.8(0.9) & 43.8(1.3) & 42.7(2.0) & 40.2(1.6) & 25.2(2.8) & 4.3(1.5) \\
\textsc{LLaDA-1.5} & pass@5 & 42.1 & 50.9(0.3) & 54.4(1.2) & 55.6(0.9) & \textbf{60.2(1.2)} & 54.3(1.8) & 22.7(1.6) \\
& pass@1 & 42.1 & \textbf{45.5(1.3)} & 44.8(1.3) & 43.0(2.0) & 41.2(1.6) & 29.1(2.6) & 7.6(1.9) \\
\textsc{Dream-v0-Instruct-7B} & pass@5 & 41.5 & 57.9(0.0) & 46.3(0.0) & 42.0(0.5) & 56.2(2.0) & 63.0(1.6) & \textbf{66.3(1.2)} \\
& pass@1 & 41.5 & \textbf{57.9(0.0)} & 45.5(0.7) & 33.9(1.5) & 38.9(1.8) & 40.3(3.0) & 38.6(3.0) \\
\textsc{Dream-Coder-v0-Instruct-7B} & pass@5 & 59.8 & \textbf{75.0(0.0)} & 70.0(0.3) & 53.8(0.8) & 48.8(1.2) & 57.6(1.3) & 70.2(0.9) \\
& pass@1 & 59.8 & \textbf{75.0(0.0)} & 69.5(0.4) & 50.5(1.0) & 39.5(1.9) & 42.3(2.7) & 48.5(2.7) \\
\textsc{DreamOn-v0-7B} & pass@5 & 45.1 & 49.5(0.3) & 47.9(0.5) & 59.6(2.3) & 65.4(1.2) & \textbf{68.3(1.6)} & 65.7(1.4) \\
& pass@1 & 45.1 & \textbf{49.4(0.1)} & 46.0(0.5) & 46.8(2.6) & 46.6(1.9) & 46.7(1.9) & 42.5(2.7) \\
\textsc{DiffuCoder-7B-cpGRPO} & pass@5 & 61.6 & 66.5(0.0) & \textbf{69.5(0.0)} & 67.4(0.7) & 67.4(1.1) & 68.4(0.9) & 68.9(0.4) \\
& pass@1 & 61.6 & 66.5(0.0) & \textbf{69.5(0.0)} & 65.8(0.8) & 61.8(1.3) & 60.2(1.5) & 59.3(1.8) \\
\midrule
\multicolumn{9}{l}{\textbf{LiveCodeBench}} \\
\textsc{LLaDA-8B-Instruct} & pass@5 & 7.6(0.0) & 7.4(0.1) & 8.9(0.2) & 10.4(0.4) & \textbf{11.7(0.4)} & 7.7(0.4) & 0.9(0.3) \\
& pass@1 & \textbf{7.6(0.0)} & 6.1(0.1) & 6.0(0.1) & 6.4(0.2) & 6.1(0.1) & 3.2(0.2) & 0.2(0.1) \\
\textsc{LLaDA-1.5} & pass@5 & 6.2(0.0) & 8.0(0.5) & 10.0(0.4) & 12.3(1.1) & \textbf{13.3(0.5)} & 8.3(1.0) & 1.5(0.0) \\
& pass@1 & 6.2(0.0) & 6.2(0.4) & 6.8(0.7) & \textbf{7.0(0.9)} & 6.5(0.9) & 3.8(0.7) & 0.4(0.2) \\
\textsc{Dream-v0-Instruct-7B} & pass@5 & 5.3(0.0) & \textbf{9.1(0.0)} & 5.3(0.0) & 6.2(0.1) & 6.7(0.8) & 0.7(0.2) & 0.1(0.1) \\
& pass@1 & 5.3(0.0) & \textbf{9.1(0.0)} & 4.7(0.1) & 3.7(0.1) & 2.6(0.2) & 0.2(0.1) & 0.0(0.0) \\
\textsc{Dream-Coder-v0-Instruct-7B} & pass@5 & 5.3(0.0) & \textbf{12.0(0.0)} & 10.3(0.0) & 8.0(0.5) & 10.6(0.6) & 11.2(0.8) & 4.3(0.7) \\
& pass@1 & 5.3(0.0) & \textbf{12.0(0.0)} & 9.9(0.2) & 7.0(0.3) & 5.5(0.8) & 4.3(0.9) & 1.4(0.7) \\
\textsc{DiffuCoder-7B-cpGRPO} & pass@5 & 6.5(0.0) & 7.3(0.0) & 7.3(0.0) & 7.0(0.1) & 7.8(0.1) & \textbf{10.6(0.7)} & 7.4(0.8) \\
& pass@1 & 6.5(0.0) & \textbf{7.3(0.0)} & 7.1(0.2) & 6.3(0.3) & 5.2(0.5) & 5.3(1.0) & 2.5(0.8) \\
\bottomrule
\end{tabular}
}
\vspace{-0.04in}
\end{table}

\cl{\textbf{Temperature affects pass@1 and pass@5 differently, with model-specific optima.} For \textsc{LLaDA-1.5} on HumanEval, pass@1 peaks at temperature 0.2 (45.5\%), whereas pass@5 peaks at 1.2 (60.2\%). For the two inspected AR baselines, pass@1 at temperature 0 is higher than at 1.2: \textsc{CodeLlama-7B-Instruct-hf} scores 41.5\% versus 38.2\%, and \textsc{Seed-Coder-8B-Instruct} scores 63.4\% versus 58.6\%.}

\cl{On LiveCodeBench, the optimal temperature varies across models and metrics. For instance, \textsc{LLaDA-8B-Instruct} achieves its best pass@5 (11.7\%) at temperature 1.2, while \textsc{Dream-Coder-v0-Instruct-7B} peaks at 0.2 (12.0\%). Across HumanEval-X, lower nonzero temperatures often improve pass@1 for the Dream and DiffuCoder models, whereas the LLaDA variants frequently peak at zero or near-zero temperature; pass@5 often favors higher temperatures.}

\begin{table}[!t]
\centering
\small
\caption{\cl{pass@5 (\%) and pass@1 (\%) across different temperatures on HumanEval-X. Values in parentheses are standard deviations over five runs.}}
\label{tab:rq2_temperature_new}

\adjustbox{max width=\linewidth}{
\begin{tabular}{@{} l l c c c c c c c @{}}
\toprule
Model & Metric & \multicolumn{7}{c}{Temperature} \\
\cmidrule(lr){3-9}
& & 0.0 & 0.2 & 0.4 & 0.8 & 1.2 & 1.6 & 2.0 \\
\midrule
\multicolumn{9}{l}{HumanEval-X (C++)} \\
\textsc{LLaDA-8B-Instruct} & pass@5 & 26.8(0.0) & 30.2(0.5) & 33.8(0.3) & 38.3(1.0) & 39.5(0.9) & 32.6(1.1) & 9.1(0.7) \\
& pass@1 & 26.8(0.0) & 26.5(0.2) & 26.9(0.3) & 27.0(0.4) & 25.8(0.3) & 16.9(0.3) & 2.6(0.2) \\
\textsc{LLaDA-1.5} & pass@5 & 28.7(0.0) & 32.2(0.3) & 34.1(1.1) & 38.4(2.6) & 42.0(0.5) & 37.7(0.8) & 14.0(0.6) \\
& pass@1 & 28.7(0.0) & 28.9(0.9) & 28.5(1.2) & 27.8(1.5) & 27.0(1.9) & 20.6(1.9) & 5.7(1.5) \\
\textsc{Dream-v0-Instruct-7B} & pass@5 & 26.2(0.0) & 40.2(0.0) & 26.7(0.3) & 29.9(0.6) & 34.3(1.4) & 13.0(0.8) & 2.6(0.7) \\
& pass@1 & 26.2(0.0) & 40.2(0.0) & 25.8(0.2) & 18.2(0.4) & 12.8(0.4) & 3.9(0.3) & 0.8(0.1) \\
\textsc{Dream-Coder-v0-Instruct-7B} & pass@5 & 30.5(0.0) & 63.4(1.9) & 54.8(0.3) & 46.2(0.8) & 54.6(0.9) & 54.1(2.0) & 28.2(2.9) \\
& pass@1 & 30.5(0.0) & 62.8(1.5) & 54.4(0.4) & 37.5(1.7) & 33.5(2.1) & 26.0(2.4) & 10.5(2.2) \\
\textsc{DiffuCoder-7B-cpGRPO} & pass@5 & 32.9(0.0) & 44.5(0.0) & 42.7(0.0) & 37.4(0.3) & 50.2(1.3) & 56.1(1.3) & 48.9(2.2) \\
& pass@1 & 32.9(0.0) & 44.5(0.0) & 42.7(0.0) & 33.2(0.9) & 34.3(1.6) & 32.2(2.4) & 23.6(3.0) \\
\midrule
\multicolumn{9}{l}{HumanEval-X (Java)} \\
\textsc{LLaDA-8B-Instruct} & pass@5 & 41.5(0.0) & 44.5(0.0) & 47.9(0.3) & 52.6(0.7) & 54.4(0.5) & 47.4(0.7) & 10.9(0.5) \\
& pass@1 & 41.5(0.0) & 40.8(0.3) & 39.8(0.3) & 38.3(0.3) & 35.1(0.4) & 22.8(0.5) & 3.9(0.3) \\
\textsc{LLaDA-1.5} & pass@5 & 36.6(0.0) & 40.7(1.1) & 44.4(1.2) & 49.8(2.4) & 53.7(0.7) & 46.5(2.3) & 17.2(2.5) \\
& pass@1 & 36.6(0.0) & 36.0(1.0) & 36.4(1.8) & 36.5(2.0) & 34.9(1.7) & 24.1(2.4) & 5.9(1.8) \\
\textsc{Dream-v0-Instruct-7B} & pass@5 & 40.9(0.0) & 53.7(0.0) & 48.2(0.0) & 42.7(0.0) & 42.0(1.2) & 16.8(0.5) & 4.8(0.5) \\
& pass@1 & 40.9(0.0) & 53.5(0.1) & 46.5(0.2) & 28.4(0.4) & 16.5(0.6) & 5.7(0.3) & 1.3(0.1) \\
\textsc{Dream-Coder-v0-Instruct-7B} & pass@5 & 41.5(0.0) & 67.7(0.0) & 63.4(0.0) & 53.2(0.3) & 57.8(2.7) & 57.6(0.5) & 32.2(1.8) \\
& pass@1 & 41.5(0.0) & 67.7(0.0) & 62.3(0.5) & 46.6(1.2) & 36.9(2.1) & 27.4(2.7) & 11.9(2.1) \\
\textsc{DiffuCoder-7B-cpGRPO} & pass@5 & 39.6(0.0) & 53.7(0.0) & 54.3(0.0) & 42.2(0.9) & 48.7(0.7) & 60.7(1.3) & 60.1(2.7) \\
& pass@1 & 39.6(0.0) & 53.7(0.0) & 54.0(0.3) & 39.0(0.8) & 35.8(1.8) & 38.6(2.3) & 29.5(2.9) \\
\midrule
\multicolumn{9}{l}{HumanEval-X (Go)} \\
\textsc{LLaDA-8B-Instruct} & pass@5 & 30.5(0.0) & 32.6(0.3) & 36.1(0.7) & 39.8(0.7) & 38.7(0.5) & 30.6(1.1) & 6.3(0.3) \\
& pass@1 & 30.5(0.0) & 27.8(0.2) & 27.9(0.3) & 25.9(0.4) & 22.6(0.4) & 13.0(0.3) & 2.4(0.2) \\
\textsc{LLaDA-1.5} & pass@5 & 26.8(0.0) & 31.1(1.0) & 34.6(0.9) & 38.0(1.7) & 41.5(2.4) & 33.8(1.4) & 10.7(2.0) \\
& pass@1 & 26.8(0.0) & 26.5(0.9) & 26.2(1.8) & 25.3(2.1) & 23.8(2.5) & 16.1(1.8) & 4.1(1.0) \\
\textsc{Dream-v0-Instruct-7B} & pass@5 & 28.0(0.0) & 37.8(0.0) & 28.7(0.0) & 30.2(0.8) & 23.0(2.6) & 11.1(0.8) & 2.9(0.5) \\
& pass@1 & 28.0(0.0) & 37.8(0.0) & 26.4(0.2) & 17.9(0.4) & 7.7(0.5) & 3.5(0.2) & 0.8(0.2) \\
\textsc{Dream-Coder-v0-Instruct-7B} & pass@5 & 29.9(0.0) & 63.4(0.0) & 52.3(1.2) & 40.6(1.0) & 48.8(1.0) & 47.9(3.1) & 27.1(3.1) \\
& pass@1 & 29.9(0.0) & 63.4(0.0) & 51.7(1.3) & 33.6(1.6) & 28.3(2.0) & 20.8(1.9) & 9.5(1.7) \\
\textsc{DiffuCoder-7B-cpGRPO} & pass@5 & 25.6(0.0) & 43.9(0.0) & 41.5(0.0) & 31.3(0.5) & 36.6(1.4) & 49.1(2.6) & 46.7(1.3) \\
& pass@1 & 25.6(0.0) & 43.9(0.0) & 41.5(0.0) & 28.7(0.8) & 24.4(1.9) & 23.6(2.2) & 19.6(2.3) \\
\midrule
\multicolumn{9}{l}{HumanEval-X (JavaScript)} \\
\textsc{LLaDA-8B-Instruct} & pass@5 & 26.2(0.0) & 30.2(0.5) & 32.6(0.5) & 35.1(0.8) & 38.4(0.7) & 32.0(0.3) & 5.0(0.5) \\
& pass@1 & 26.2(0.0) & 25.6(0.3) & 25.9(0.3) & 24.1(0.4) & 21.5(0.5) & 12.5(0.5) & 1.2(0.2) \\
\textsc{LLaDA-1.5} & pass@5 & 32.9(0.0) & 36.0(0.7) & 39.0(0.7) & 43.0(1.9) & 44.5(1.9) & 37.6(1.8) & 12.4(0.5) \\
& pass@1 & 32.9(0.0) & 33.4(0.9) & 32.7(1.3) & 30.4(2.1) & 27.4(2.7) & 19.0(2.4) & 4.1(1.2) \\
\textsc{Dream-v0-Instruct-7B} & pass@5 & 49.4(0.0) & 56.1(0.0) & 46.6(0.3) & 37.0(1.4) & 45.7(1.8) & 20.1(1.0) & 6.1(0.6) \\
& pass@1 & 49.4(0.0) & 55.4(0.1) & 45.3(0.1) & 25.7(0.4) & 17.9(0.5) & 6.8(0.3) & 1.7(0.2) \\
\textsc{Dream-Coder-v0-Instruct-7B} & pass@5 & 41.5(0.0) & 51.8(0.0) & 56.7(0.0) & 49.9(1.6) & 64.0(2.7) & 59.5(2.3) & 34.3(2.7) \\
& pass@1 & 41.5(0.0) & 51.8(0.0) & 56.6(0.3) & 43.1(1.2) & 40.5(1.3) & 28.6(3.3) & 12.8(2.1) \\
\textsc{DiffuCoder-7B-cpGRPO} & pass@5 & 47.6(0.0) & 56.7(0.0) & 55.5(0.0) & 53.8(0.8) & 61.0(1.7) & 65.4(2.2) & 53.4(2.5) \\
& pass@1 & 47.6(0.0) & 56.6(0.3) & 55.5(0.0) & 47.0(1.4) & 44.2(2.4) & 37.5(2.3) & 18.9(2.1) \\
\bottomrule
\end{tabular}
}
\vspace{-0.04in}
\end{table}

\cl{\textbf{Moderate randomness can improve pass@k for some diffusion models, but the benefit depends on the model, benchmark, and value of $k$; our experiments do not isolate the cause of these differences.}}

\begin{boxK}
\small \faIcon{pencil-alt} \textbf{Answer to \cl{RQ3}:}
\cc
\cl{Across the evaluated settings, increasing diffusion steps generally improves performance, and confidence-based remasking consistently outperforms random selection. Generation length, block length, and temperature have model- and task-dependent optima, while very long blocks often reduce effectiveness. These findings support configuration-specific tuning rather than a single setting shared by all diffusion LLMs.}
\end{boxK}

\FloatBarrier
\subsection{\cl{RQ4: Impact of Diffusion Steps and Generation Length on Efficiency}}
\label{sec:rq3}

\textbf{Motivation.}
Efficiency is a critical consideration in code generation, where both real-time responsiveness and low inference cost are often required. \cl{A potential advantage of diffusion LLMs is their support for parallel decoding. In this section, we systematically evaluate how diffusion steps and generation length affect efficiency in code generation.}

\cl{These two factors directly determine the number of denoising passes and the sequence length processed per pass. In contrast, the other factors studied in RQ3 primarily affect effectiveness and are expected to have comparatively small influence on efficiency under the evaluated setup.}

\subsubsection{Impact of Diffusion Steps}
\label{sec:rq3_step}
\cl{The experimental setting follows Section~\ref{sec:rq2_step}. We evaluate five open-source diffusion LLMs on HumanEval and MBPP and two representative diffusion LLMs on LiveCodeBench and HumanEval-X (C++), with \textsc{CodeLlama-7B-Instruct-hf} as the AR baseline.}

\cl{The evaluated results, including pass@1 where reported for trade-off analysis, are shown in Table~\ref{tab:rq3_step_eff} for HumanEval and MBPP and in Table~\ref{tab:rq3_step_eff_new} for LiveCodeBench and HumanEval-X (C++).}

\begin{table}[!t]
\centering
\small
\caption{\cl{pass@1 on HumanEval, and throughput, FLOPs/token, and average decoding time on HumanEval and MBPP, across different diffusion steps.}}
\label{tab:rq3_step_eff}
\adjustbox{max width=\linewidth}{
\begin{tabular}{@{} l c c c c c c c c @{}}
\toprule
\multirow{2}{*}{\textbf{Model}} & \multirow{2}{*}{\textbf{Steps}} & \multicolumn{4}{c}{\textbf{HumanEval}} & \multicolumn{3}{c}{\textbf{MBPP}} \\
\cmidrule(lr){3-6} \cmidrule(lr){7-9}
& & \textbf{pass@1} & \textbf{token/s} & \textbf{FLOPs/token} & \textbf{s/q} & \textbf{token/s} & \textbf{FLOPs/token} & \textbf{s/q} \\
\midrule
\textsc{Dream-Coder-v0-Instruct-7B} & 512 & 59.8 & 12.2 & $1.17{\times}10^{13}$ & 41.9 & 14.2 & $9.41{\times}10^{12}$ & 36.0 \\
 & 256 & 57.9 & 24.4 & $5.87{\times}10^{12}$ & 21.0 & 27.9 & $4.71{\times}10^{12}$ & 18.4 \\
 & 128 & 55.5 & 48.8 & $2.94{\times}10^{12}$ & 10.5 & 57.1 & $2.34{\times}10^{12}$ & 9.0 \\
 & 64 & 40.9 & 97.6 & $1.47{\times}10^{12}$ & 5.2 & 113.2 & $1.18{\times}10^{12}$ & 4.5 \\
 & 32 & 23.2 & 194.4 & $7.34{\times}10^{11}$ & 2.6 & 225.9 & $5.88{\times}10^{11}$ & 2.3 \\
 & 16 & 14.6 & 386.3 & $3.67{\times}10^{11}$ & 1.3 & 448.6 & $2.95{\times}10^{11}$ & 1.1 \\
 & 8 & 12.2 & 765.3 & $1.83{\times}10^{11}$ & 0.7 & 883.9 & $1.47{\times}10^{11}$ & 0.6 \\
\midrule
\textsc{DiffuCoder-7B-cpGRPO} & 512 & 61.6 & 13.0 & $1.17{\times}10^{13}$ & 39.5 & 15.1 & $9.59{\times}10^{12}$ & 34.0 \\
 & 256 & 62.2 & 25.7 & $5.88{\times}10^{12}$ & 19.9 & 30.1 & $4.80{\times}10^{12}$ & 17.0 \\
 & 128 & 51.8 & 51.4 & $2.94{\times}10^{12}$ & 10.0 & 57.5 & $2.40{\times}10^{12}$ & 8.9 \\
 & 64 & 41.5 & 102.7 & $1.47{\times}10^{12}$ & 5.0 & 120.4 & $1.20{\times}10^{12}$ & 4.3 \\
 & 32 & 38.4 & 204.6 & $7.32{\times}10^{11}$ & 2.5 & 239.1 & $6.00{\times}10^{11}$ & 2.1 \\
 & 16 & 33.5 & 408.2 & $3.67{\times}10^{11}$ & 1.3 & 473.9 & $3.01{\times}10^{11}$ & 1.1 \\
 & 8 & 28.7 & 816.4 & $1.84{\times}10^{11}$ & 0.6 & 927.3 & $1.50{\times}10^{11}$ & 0.6 \\
\midrule
\textsc{LLaDA-1.5} & 512 & 42.1 & 11.3 & $1.29{\times}10^{13}$ & 45.5 & 15.1 & $1.02{\times}10^{13}$ & 34.0 \\
 & 256 & 34.8 & 22.5 & $6.46{\times}10^{12}$ & 22.7 & 28.5 & $5.14{\times}10^{12}$ & 18.0 \\
 & 128 & 18.9 & 44.9 & $3.22{\times}10^{12}$ & 11.4 & 55.5 & $2.56{\times}10^{12}$ & 9.2 \\
 & 64 & 9.1 & 90.0 & $1.62{\times}10^{12}$ & 5.7 & 120.5 & $1.28{\times}10^{12}$ & 4.2 \\
 & 32 & 4.9 & 181.2 & $8.07{\times}10^{11}$ & 2.8 & 240.8 & $6.39{\times}10^{11}$ & 2.1 \\
 & 16 & 1.8 & 362.1 & $4.04{\times}10^{11}$ & 1.4 & 480.0 & $3.20{\times}10^{11}$ & 1.1 \\
 & 8 & 1.2 & 816.4 & $2.01{\times}10^{11}$ & 0.6 & 959.6 & $1.60{\times}10^{11}$ & 0.5 \\
\midrule
\textsc{LLaDA-8B-Instruct} & 512 & 45.7 & 9.6 & $1.29{\times}10^{13}$ & 53.4 & 12.7 & $1.02{\times}10^{13}$ & 40.2 \\
 & 256 & 34.1 & 19.3 & $6.46{\times}10^{12}$ & 26.6 & 25.0 & $5.12{\times}10^{12}$ & 20.5 \\
 & 128 & 22.6 & 38.4 & $3.22{\times}10^{12}$ & 13.3 & 51.0 & $2.56{\times}10^{12}$ & 10.0 \\
 & 64 & 7.9 & 78.2 & $1.62{\times}10^{12}$ & 6.5 & 100.5 & $1.28{\times}10^{12}$ & 5.1 \\
 & 32 & 3.7 & 156.8 & $8.07{\times}10^{11}$ & 3.3 & 205.3 & $6.39{\times}10^{11}$ & 2.5 \\
 & 16 & 3.7 & 312.1 & $4.04{\times}10^{11}$ & 1.6 & 410.7 & $3.20{\times}10^{11}$ & 1.2 \\
 & 8 & 0.6 & 615.0 & $2.01{\times}10^{11}$ & 0.8 & 812.6 & $1.60{\times}10^{11}$ & 0.6 \\
\midrule
\textsc{Dream-v0-Instruct-7B} & 512 & 41.5 & 12.9 & $1.19{\times}10^{13}$ & 39.8 & 14.5 & $9.47{\times}10^{12}$ & 35.3 \\
 & 256 & 34.8 & 25.8 & $6.15{\times}10^{12}$ & 19.9 & 29.5 & $4.73{\times}10^{12}$ & 17.4 \\
 & 128 & 34.8 & 50.8 & $2.97{\times}10^{12}$ & 10.1 & 60.8 & $2.36{\times}10^{12}$ & 8.4 \\
 & 64 & 19.5 & 103.0 & $1.48{\times}10^{12}$ & 5.0 & 119.5 & $1.18{\times}10^{12}$ & 4.3 \\
 & 32 & 6.7 & 204.8 & $7.44{\times}10^{11}$ & 2.5 & 241.3 & $5.92{\times}10^{11}$ & 2.1 \\
 & 16 & 1.8 & 413.1 & $3.71{\times}10^{11}$ & 1.2 & 479.9 & $2.95{\times}10^{11}$ & 1.1 \\
 & 8 & 1.2 & 810.5 & $1.85{\times}10^{11}$ & 0.6 & 939.7 & $1.48{\times}10^{11}$ & 0.5 \\
\midrule
\multicolumn{9}{l}{\textbf{AR Baseline (\textsc{CodeLlama-7B-Instruct-hf})}} \\
 & w/ KV cache & 41.5 & 40.0 & $4.63{\times}10^{10}$ & 4.0 & 39.2 & $2.18{\times}10^{10}$ & 7.5 \\
 & w/o KV cache & 41.5 & 18.1 & $7.66{\times}10^{12}$ & 8.3 & 24.5 & $5.12{\times}10^{12}$ & 11.7 \\
\bottomrule
\end{tabular}
}
\end{table}

\cl{\textbf{Reducing the number of diffusion steps decreases computational cost and improves time efficiency.}}
Fewer steps lead to fewer forward passes, which reduces FLOPs/token and increases throughput.
\cl{This trend also leads to a clear reduction in average decoding time.}
For example, reducing steps from 512 to 256 approximately halves the FLOPs/token of \textsc{DiffuCoder-7B-cpGRPO}, from $1.17\times 10^{13}$ to $5.88\times 10^{12}$.
Its average decoding time on HumanEval drops from 39.5s/question to 19.9s/question.
Similarly, reducing steps from 16 to 8 increases the throughput of \textsc{Dream-Coder-v0-Instruct-7B} from 386.3 to 765.3 tokens/s, and its average decoding time decreases from 1.3s/question to 0.7s/question.

\begin{table}[!t]
\centering
\small
\caption{\cl{pass@1 (\%), throughput (token/s), FLOPs/token, and average decoding time (s/question) across different diffusion steps on LiveCodeBench and HumanEval-X (C++).}}
\label{tab:rq3_step_eff_new}
\adjustbox{max width=\linewidth}{
\begin{tabular}{@{} l c c c c c c c c c @{}}
\toprule
\multirow{2}{*}{\textbf{Model}} & \multirow{2}{*}{\textbf{Steps}} & \multicolumn{4}{c}{\textbf{LiveCodeBench}} & \multicolumn{4}{c}{\textbf{HumanEval-X (C++)}} \\
\cmidrule(lr){3-6} \cmidrule(lr){7-10}
& & \textbf{pass@1} & \textbf{token/s} & \textbf{FLOPs/token} & \textbf{s/q} & \textbf{pass@1} & \textbf{token/s} & \textbf{FLOPs/token} & \textbf{s/q} \\
\midrule
\textsc{LLaDA-8B-Instruct} & 512 & 6.2 & 9.2 & $1.72{\times}10^{13}$ & 55.7 & 25.6 & 11.3 & $1.32{\times}10^{13}$ & 45.3\\

 & 256 & 5.3 & 18.4 & $8.62{\times}10^{12}$ & 27.9 & 19.5 & 22.6 & $6.62{\times}10^{12}$ & 22.6\\

 & 128 & 0.6 & 36.7 & $4.31{\times}10^{12}$ & 13.9 & 9.8 & 45.4 & $3.31{\times}10^{12}$ & 11.3\\

 & 64 & 0.0 & 73.5 & $2.16{\times}10^{12}$ & 7.0 & 1.8 & 90.5 & $1.65{\times}10^{12}$ & 5.7\\

 & 32 & 0.0 & 147.2 & $1.08{\times}10^{12}$ & 3.5 & 1.2 & 181.2 & $8.27{\times}10^{11}$ & 2.8\\

 & 16 & 0.0 & 294.2 & $5.39{\times}10^{11}$ & 1.7 & 0.0 & 362.0 & $4.14{\times}10^{11}$ & 1.4\\

 & 8 & 0.0 & 587.7 & $2.70{\times}10^{11}$ & 0.9 & 0.0 & 723.8 & $2.07{\times}10^{11}$ & 0.7\\

\midrule
\textsc{Dream-v0-Instruct-7B} & 512 & 1.8 & 10.3 & $1.56{\times}10^{13}$ & 49.7 & 15.9 & 12.8 & $1.21{\times}10^{13}$ & 39.9\\

 & 256 & 2.9 & 20.6 & $7.78{\times}10^{12}$ & 24.9 & 21.3 & 25.7 & $6.04{\times}10^{12}$ & 19.9\\

 & 128 & 2.1 & 41.1 & $3.89{\times}10^{12}$ & 12.5 & 15.9 & 51.5 & $3.02{\times}10^{12}$ & 9.9\\

 & 64 & 1.5 & 82.3 & $1.94{\times}10^{12}$ & 6.2 & 8.5 & 102.9 & $1.51{\times}10^{12}$ & 5.0\\

 & 32 & 0.3 & 164.6 & $9.72{\times}10^{11}$ & 3.1 & 2.4 & 205.4 & $7.55{\times}10^{11}$ & 2.5\\

 & 16 & 0.0 & 328.5 & $4.86{\times}10^{11}$ & 1.6 & 1.8 & 408.2 & $3.77{\times}10^{11}$ & 1.3\\

 & 8 & 0.0 & 654.4 & $2.43{\times}10^{11}$ & 0.8 & 1.2 & 813.7 & $1.89{\times}10^{11}$ & 0.6\\

\midrule
\multicolumn{10}{l}{\textbf{AR Baseline (\textsc{CodeLlama-7B-Instruct-hf})}} \\
 & w/ KV cache & 9.2 & 31.9 & $6.07{\times}10^{10}$ & 5.0 & 30.0 & 39.7 & $4.71{\times}10^{10}$ & 4.0\\

 & w/o KV cache & 9.2 & 14.5 & $1.00{\times}10^{13}$ & 10.4 & 30.0 & 18.0 & $7.79{\times}10^{12}$ & 8.3\\

\bottomrule
\end{tabular}
}
\end{table}

\cl{On LiveCodeBench and HumanEval-X, the efficiency trends are consistent with those on HumanEval and MBPP. For instance, \textsc{LLaDA-8B-Instruct} on LiveCodeBench sees its throughput increase from 9.2 tokens/s at 512 steps to 587.7 tokens/s at 8 steps, while its average decoding time drops from 55.7s/question to 0.9s/question. On HumanEval-X (C++), the same model achieves 11.3 tokens/s at 512 steps and 723.8 tokens/s at 8 steps.}

\textbf{There exists a clear trade-off between efficiency and effectiveness, and average decoding time makes this trade-off more explicit when comparing against AR baselines.}
As shown in Table~\ref{tab:rq3_step_eff}, when the number of steps in \textsc{DiffuCoder-7B-cpGRPO} is decreased from 512 to 8, its throughput on HumanEval rises from 13.0 to 816.4 tokens/s, while its pass@1 drops from 61.6\% to 28.7\%.
\cl{The pass@1--average-decoding-time trade-off can be read directly from the table. For \textsc{DiffuCoder-7B-cpGRPO}: at 512 steps, pass@1 = 61.6\%, decoding time = 39.5s/question; at 256 steps, pass@1 = 62.2\%, decoding time = 19.9s/question; at 128 steps, pass@1 = 51.8\%, decoding time = 10.0s/question; at 64 steps, pass@1 = 41.5\%, decoding time = 5.0s/question; at 32 steps, pass@1 = 38.4\%, decoding time = 2.5s/question.}
To catch up with \textsc{CodeLlama-7B-Instruct-hf} \textbf{without KV cache} in average decoding time on HumanEval, \textsc{DiffuCoder-7B-cpGRPO} needs to reduce diffusion steps to around \textbf{64}.
At 64 steps, its average decoding time becomes 5.0s/question \cl{with pass@1 = 41.5\%}, compared with 8.3s/question for \textsc{CodeLlama-7B-Instruct-hf} without KV cache.
To catch up with \textsc{CodeLlama-7B-Instruct-hf} \textbf{with KV cache}, diffusion steps typically need to be reduced to around \textbf{32}, where \textsc{DiffuCoder-7B-cpGRPO} reaches 2.5s/question \cl{with pass@1 = 38.4\%}, compared with 4.0s/question for \textsc{CodeLlama-7B-Instruct-hf} with KV cache.
Nevertheless, we find that an appropriate choice of diffusion steps can still balance efficiency and effectiveness in many settings.
\cl{At 256 steps, \textsc{DiffuCoder-7B-cpGRPO} preserves its 512-step pass@1 (62.2\% vs.\ 61.6\%) while approximately halving decoding time; at 128 steps, decoding time falls further but pass@1 decreases to 51.8\%.}

\cl{\textbf{Some evaluated diffusion configurations surpass the AR baseline without KV cache on FLOPs/token and throughput; comparison with KV-cached AR decoding is less favorable.}}
On HumanEval, at \textbf{256} steps, \textsc{DiffuCoder-7B-cpGRPO} achieves \textbf{25.7 tokens/s} and \textbf{$5.88\times 10^{12}$ FLOPs/token}.
This surpasses \textsc{CodeLlama-7B-Instruct-hf} \textbf{without KV cache}, which achieves \textbf{18.1 tokens/s} and \textbf{$7.66\times 10^{12}$ FLOPs/token}.
A similar pattern holds for \textsc{Dream-Coder-v0-Instruct-7B}, which achieves 24.4 tokens/s and $5.87\times 10^{12}$ FLOPs/token at 256 steps.
When KV cache is enabled, \textsc{CodeLlama-7B-Instruct-hf} achieves 40.0 tokens/s and $4.63\times 10^{10}$ FLOPs/token.
In this case, diffusion models typically surpass the AR baseline on throughput only when steps are reduced to around \textbf{128}, but they do not surpass it on FLOPs/token across the evaluated step range.
For end-to-end efficiency measured by average decoding time, the cross-over is more stringent.
\cl{As shown in Table~\ref{tab:rq3_step_eff},} diffusion models become faster than AR without KV cache at around \textbf{64} steps, and faster than AR with KV cache at around \textbf{32} steps.

\cl{\textbf{We note that diffusion LLMs currently lack native KV cache support, which represents a practical limitation.} Unlike AR models that can cache key-value pairs of previously generated tokens, the bidirectional attention and iterative denoising of diffusion LLMs make reusing cached states across steps non-trivial. Each diffusion step attends over the entire sequence, and the mask pattern changes between steps, complicating incremental caching. While recent work has begun exploring KV cache mechanisms for diffusion LLMs~\cite{fastdllm, liu2025dllm, wang2025diffusion}, these techniques are still at an early stage and are not yet widely adopted. We therefore report AR efficiency both with and without KV cache, allowing a more complete picture of the current efficiency landscape.}

\textbf{A notable observation is that enabling KV cache dramatically reduces FLOPs/token for AR decoding, but the improvements in throughput and average decoding time are much smaller.}
\cl{As shown in Table~\ref{tab:rq3_step_eff}}, enabling KV cache reduces the FLOPs/token of \textsc{CodeLlama-7B-Instruct-hf} from $7.66\times 10^{12}$ to $4.63\times 10^{10}$.
However, \cl{the same table} shows that throughput increases from 18.1 to 40.0 tokens/s, and average decoding time decreases from 8.3s/question to 4.0s/question.
This gap suggests that FLOPs/token alone may overestimate practical speedups under KV-cached decoding.

\FloatBarrier

\subsubsection{Impact of Generation Length}
\label{sec:rq3_length}

\cl{The experimental setting follows Section~\ref{sec:rq2_length}. We evaluate five open-source diffusion LLMs on HumanEval and MBPP and two representative diffusion LLMs on LiveCodeBench and HumanEval-X (C++), with \textsc{CodeLlama-7B-Instruct-hf} as the AR baseline.}

\cl{The evaluated results are shown for HumanEval and MBPP in Table~\ref{tab:rq3_length_eff} and for LiveCodeBench and HumanEval-X (C++) in Table~\ref{tab:rq3_length_eff_new}.}

\begin{table}[!t]
\centering
\small
\caption{\cl{pass@1 on HumanEval, and throughput, FLOPs/token, and average decoding time on HumanEval and MBPP, across different generation lengths.}}
\label{tab:rq3_length_eff}
\adjustbox{max width=\linewidth}{
\begin{tabular}{@{} l c c c c c c c c @{}}
\toprule
\multirow{2}{*}{\textbf{Model}} & \multirow{2}{*}{\textbf{Length}} & \multicolumn{4}{c}{\textbf{HumanEval}} & \multicolumn{3}{c}{\textbf{MBPP}} \\
\cmidrule(lr){3-6} \cmidrule(lr){7-9}
& & \textbf{pass@1} & \textbf{token/s} & \textbf{FLOPs/token} & \textbf{s/q} & \textbf{token/s} & \textbf{FLOPs/token} & \textbf{s/q} \\
\midrule
\textsc{Dream-Coder-v0-Instruct-7B} & 2048 & 52.4 & 4.3 & $3.54{\times}10^{13}$ & 480.6 & 4.5 & $3.29{\times}10^{13}$ & 452.3\\

 & 1024 & 56.1 & 7.6 & $1.94{\times}10^{13}$ & 134.9 & 8.3 & $1.70{\times}10^{13}$ & 123.2\\

 & 512 & 59.8 & 12.2 & $1.17{\times}10^{13}$ & 41.9 & 14.2 & $9.41{\times}10^{12}$ & 36.0\\

 & 256 & 59.1 & 16.7 & $7.98{\times}10^{12}$ & 15.3 & 20.8 & $5.66{\times}10^{12}$ & 12.3\\

 & 128 & 64.0 & 21.0 & $6.12{\times}10^{12}$ & 6.1 & 28.7 & $3.84{\times}10^{12}$ & 4.5\\

 & 64 & 40.9 & 23.4 & $5.20{\times}10^{12}$ & 2.7 & 30.4 & $2.92{\times}10^{12}$ & 2.1\\

\midrule
\textsc{DiffuCoder-7B-cpGRPO} & 2048 & 61.0 & 4.7 & $3.54{\times}10^{13}$ & 440.3 & 4.9 & $3.31{\times}10^{13}$ & 416.8\\

 & 1024 & 61.0 & 8.2 & $1.94{\times}10^{13}$ & 125.2 & 8.9 & $1.72{\times}10^{13}$ & 114.7\\

 & 512 & 61.6 & 13.0 & $1.17{\times}10^{13}$ & 39.5 & 15.1 & $9.59{\times}10^{12}$ & 34.0\\

 & 256 & 61.6 & 17.4 & $7.97{\times}10^{12}$ & 14.7 & 21.9 & $5.86{\times}10^{12}$ & 11.7\\

 & 128 & 65.2 & 21.3 & $6.13{\times}10^{12}$ & 6.0 & 27.2 & $4.02{\times}10^{12}$ & 4.7\\

 & 64 & 54.9 & 23.8 & $5.20{\times}10^{12}$ & 2.7 & 31.1 & $3.11{\times}10^{12}$ & 2.1\\

\midrule
\textsc{LLaDA-1.5} & 2048 & 45.1 & 3.7 & $3.85{\times}10^{13}$ & 546.2 & 4.0 & $3.56{\times}10^{13}$ & 510.3\\

 & 1024 & 45.1 & 6.4 & $2.12{\times}10^{13}$ & 160.4 & 7.5 & $1.84{\times}10^{13}$ & 135.7\\

 & 512 & 42.1 & 11.3 & $1.29{\times}10^{13}$ & 45.5 & 15.1 & $1.02{\times}10^{13}$ & 34.0\\

 & 256 & 45.5 & 15.9 & $8.91{\times}10^{12}$ & 16.1 & 22.2 & $6.25{\times}10^{12}$ & 11.5\\

 & 128 & 42.1 & 19.7 & $6.90{\times}10^{12}$ & 6.5 & 28.7 & $4.28{\times}10^{12}$ & 4.5\\

 & 64 & 34.8 & 22.1 & $5.92{\times}10^{12}$ & 2.9 & 32.8 & $3.30{\times}10^{12}$ & 2.0\\

\midrule
\textsc{LLaDA-8B-Instruct} & 2048 & 45.7 & 3.8 & $3.85{\times}10^{13}$ & 537.9 & 4.0 & $3.55{\times}10^{13}$ & 510.3\\

 & 1024 & 47.0 & 6.5 & $2.12{\times}10^{13}$ & 158.1 & 7.4 & $1.84{\times}10^{13}$ & 137.8\\

 & 512 & 45.7 & 9.5 & $1.29{\times}10^{13}$ & 54.0 & 12.5 & $1.02{\times}10^{13}$ & 41.0\\

 & 256 & 42.7 & 13.4 & $8.91{\times}10^{12}$ & 19.1 & 19.0 & $6.25{\times}10^{12}$ & 13.5\\

 & 128 & 39.6 & 17.0 & $6.90{\times}10^{12}$ & 7.5 & 25.2 & $4.28{\times}10^{12}$ & 5.1\\

 & 64 & 30.5 & 19.3 & $5.92{\times}10^{12}$ & 3.3 & 28.1 & $3.30{\times}10^{12}$ & 2.3\\

\midrule
\textsc{Dream-v0-Instruct-7B} & 2048 & 27.4 & 4.6 & $3.55{\times}10^{13}$ & 444.4 & 4.9 & $3.29{\times}10^{13}$ & 417.9\\

 & 1024 & 29.9 & 8.0 & $1.95{\times}10^{13}$ & 127.8 & 9.0 & $1.71{\times}10^{13}$ & 113.6\\

 & 512 & \cl{41.5} & 12.9 & $1.19{\times}10^{13}$ & 39.8 & 14.5 & $9.47{\times}10^{12}$ & 35.3\\

 & 256 & 32.3 & 17.7 & $8.09{\times}10^{12}$ & 14.4 & 21.9 & $5.74{\times}10^{12}$ & 11.7\\

 & 128 & 34.8 & 21.2 & $6.24{\times}10^{12}$ & 6.0 & 27.2 & $3.89{\times}10^{12}$ & 4.7\\

 & 64 & 43.9 & 23.6 & $5.31{\times}10^{12}$ & 2.7 & 29.2 & $2.97{\times}10^{12}$ & 2.2\\

\midrule
\multicolumn{9}{l}{\textbf{AR Baseline (\textsc{CodeLlama-7B-Instruct-hf})}} \\
 & w/ KV cache & 41.5 & 40.0 & $4.63{\times}10^{10}$ & 4.0 & 39.2 & $2.18{\times}10^{10}$ & 7.5\\

 & w/o KV cache & 41.5 & 18.1 & $7.66{\times}10^{12}$ & 8.3 & 24.5 & $5.12{\times}10^{12}$ & 11.7\\

\bottomrule
\end{tabular}
}
\end{table}

\cl{\textbf{Longer generation lengths substantially reduce efficiency and markedly increase average decoding time.}}
For example, when the generation length increases from 64 to 2048, the FLOPs per token of \textsc{DiffuCoder-7B-cpGRPO} on MBPP increase from $3.11 \times 10^{12}$ to $3.31 \times 10^{13}$, resulting in a more than $10\times$ increase in computational cost and substantially reduced computational efficiency.
This increase in cost also impacts time efficiency, with throughput dropping from 31.1 tokens/s to \cl{4.9 tokens/s}.
Meanwhile, average decoding time increases sharply as the generation length grows.
\cl{On HumanEval, the average decoding time of \textsc{DiffuCoder-7B-cpGRPO} increases from 2.7s/question at length 64 to 440.3s/question at length 2048.}
This slowdown occurs because current diffusion LLMs adopt bidirectional attention, where all tokens participate in attention computation during each forward pass, amplifying the computational cost.
\cl{Longer fixed canvases can also increase trailing \texttt{EOS} decoding overhead.}

\begin{table}[!t]
\centering
\small
\caption{\cl{pass@1 (\%), throughput (token/s), FLOPs/token, and average decoding time (s/question) across different generation lengths on LiveCodeBench and HumanEval-X (C++).}}
\label{tab:rq3_length_eff_new}
\adjustbox{max width=\linewidth}{
\begin{tabular}{@{} l c c c c c c c c c @{}}
\toprule
\multirow{2}{*}{\textbf{Model}} & \multirow{2}{*}{\textbf{Length}} & \multicolumn{4}{c}{\textbf{LiveCodeBench}} & \multicolumn{4}{c}{\textbf{HumanEval-X (C++)}} \\
\cmidrule(lr){3-6} \cmidrule(lr){7-10}
& & \textbf{pass@1} & \textbf{token/s} & \textbf{FLOPs/token} & \textbf{s/q} & \textbf{pass@1} & \textbf{token/s} & \textbf{FLOPs/token} & \textbf{s/q} \\
\midrule
\textsc{LLaDA-8B-Instruct} & 2048 & 5.6 & 4.0 & $4.68{\times}10^{13}$ & 511.0 & -- & -- & -- & --\\

 & 1024 & 4.7 & 6.5 & $2.57{\times}10^{13}$ & 158.5 & 28.7 & 7.8 & $2.15{\times}10^{13}$ & 131.5\\

 & 512 & 6.2 & 9.0 & $1.72{\times}10^{13}$ & 56.6 & 25.6 & 11.4 & $1.32{\times}10^{13}$ & 45.0\\

 & 256 & 7.9 & 11.4 & $1.31{\times}10^{13}$ & 22.5 & 29.3 & 16.0 & $9.20{\times}10^{12}$ & 16.0\\

 & 128 & 3.8 & 13.2 & $1.11{\times}10^{13}$ & 9.7 & 25.0 & 19.6 & $7.21{\times}10^{12}$ & 6.5\\

 & 64 & 3.8 & 14.4 & $1.01{\times}10^{13}$ & 4.4 & 20.1 & 21.8 & $6.22{\times}10^{12}$ & 2.9\\

\midrule
\textsc{Dream-v0-Instruct-7B} & 2048 & 1.2 & 4.3 & $4.24{\times}10^{13}$ & 476.5 & -- & -- & -- & --\\

 & 1024 & 1.2 & 7.0 & $2.33{\times}10^{13}$ & 145.6 & 11.6 & 8.0 & $1.98{\times}10^{13}$ & 127.4\\

 & 512 & 1.8 & 10.3 & $1.56{\times}10^{13}$ & 49.8 & 15.2 & 12.6 & $1.21{\times}10^{13}$ & 40.5\\

 & 256 & 2.4 & 13.5 & $1.17{\times}10^{13}$ & 19.0 & 14.6 & 17.4 & $8.31{\times}10^{12}$ & 14.7\\

 & 128 & 3.8 & 15.7 & $9.86{\times}10^{12}$ & 8.1 & 23.8 & 21.0 & $6.45{\times}10^{12}$ & 6.1\\

 & 64 & 3.2 & 17.2 & $8.92{\times}10^{12}$ & 3.7 & 23.2 & 23.4 & $5.52{\times}10^{12}$ & 2.7\\

\midrule
\multicolumn{10}{l}{\textbf{AR Baseline (\textsc{CodeLlama-7B-Instruct-hf})}} \\
 & w/ KV cache & 9.2 & 31.9 & $6.07{\times}10^{10}$ & 5.0 & 30.0 & 39.7 & $4.71{\times}10^{10}$ & 4.0\\

 & w/o KV cache & 9.2 & 14.5 & $1.00{\times}10^{13}$ & 10.4 & 30.0 & 18.0 & $7.79{\times}10^{12}$ & 8.3\\

\bottomrule
\end{tabular}
}
\end{table}

\cl{On LiveCodeBench, the same trend holds: \textsc{LLaDA-8B-Instruct} sees its throughput decrease from 14.4 tokens/s at length 64 to 4.0 tokens/s at length 2048, while its average decoding time increases from 4.4s/question to 511.0s/question. On HumanEval-X, \textsc{LLaDA-8B-Instruct} throughput drops from 21.8 tokens/s at length 64 to 7.8 tokens/s at length 1024.}

This suggests that for practical code generation tasks it is beneficial to select a generation length that is as short as possible while still sufficient to produce complete and functional code.
It also motivates future improvements such as incorporating early stopping criteria into the diffusion process or adopting adaptive generation length control.

\begin{boxK}
\small \faIcon{pencil-alt} \textbf{Answer to \cl{RQ4}:}
\cc
\cl{For the two systematically evaluated factors, fewer diffusion steps and shorter generation lengths reduce computational cost and decoding time, but can trade off effectiveness. Longer fixed canvases can also add trailing \texttt{EOS} overhead.}
\end{boxK}

\FloatBarrier

\section{Discussion}
\label{sec:discussion}

\subsection{Future Work}
\label{sec:future}
Based on our study results and prior work, we outline several promising directions for future research on diffusion LLMs for code generation.  

\textbf{Integration of AR and Diffusion LLMs.}  
\cl{Our experiments in Section~\ref{sec:rq1} show complementary problem coverage between the selected diffusion and AR model groups; they do not establish that this complementarity is inherent to either paradigm.} This complementarity suggests that future research could explore adaptive frameworks that dynamically select the most suitable paradigm according to the characteristics of a task. Another possibility is to design collaborative systems in which AR models verify or refine the outputs of diffusion models in parallel, thereby improving both accuracy and reliability. Such hybrid approaches may also provide new insights into the broader question of how different generation paradigms can be integrated to achieve stronger overall performance.

\textbf{Structure-Aware Remasking Strategies.} 
As discussed in Section~\ref{sec:rq2_remask}, existing remasking strategies mainly include random and low-confidence remasking. While low-confidence remasking has \cl{often been found effective} in natural language tasks, code generation introduces unique structural information that can be exploited. Future strategies could leverage abstract syntax trees, control-flow graphs, or data-flow dependencies to guide the remasking process. For example, a strategy that prioritizes generating function signatures before their implementations may better reflect natural programming workflows. By incorporating structural signals, such approaches have the potential to \cl{improve} both syntactic correctness and semantic coherence.  

\textbf{Adaptive generation length.}  
Existing diffusion LLMs require a predefined generation length, which is not always optimal. \cl{As shown in Sections~\ref{sec:rq2_length} and~\ref{sec:rq3_length}, overly short lengths can prevent complete programs, whereas longer fixed canvases increase decoding cost and reduce effectiveness in some configurations; Figure~\ref{fig:case_length} illustrates one case in which the additional output is redundant commentary.} \cl{Because the required output length varies substantially across programming tasks, part of this behavior may also reflect a mismatch between a fixed generation budget and task-specific solution lengths.} Future research could investigate adaptive methods that estimate the appropriate generation length during inference based on task complexity, model confidence, or intermediate validation signals. Such adaptive mechanisms would bring diffusion LLMs closer to practical deployment, where flexibility and efficiency are equally important.

\textbf{Code completion.}  
Diffusion LLMs support in-place prompting, where the model is given a prefix, a suffix, and masked segments in between, and is tasked with filling in the masked portions. This capability makes them \cl{naturally aligned with} code completion \cl{and infilling scenarios}. Exploring how to adapt diffusion LLMs for code completion tasks presents a promising research direction. This would align closely with how programmers actually work and has the potential to integrate automated generation more seamlessly into human-in-the-loop development.  

\textbf{KV cache for diffusion LLMs.}  
Research on KV cache for diffusion LLMs is still at an early stage~\cite{fastdllm, liu2025dllm, wang2025diffusion}, which currently \cl{makes their practical efficiency less mature than AR decoding with standard KV cache}. Effective caching mechanisms could \cl{reduce} computational overhead, making diffusion models more practical for repository-level tasks. Investigating how to design KV cache structures that align with the iterative denoising process of diffusion models is a compelling direction, as it would not only improve efficiency but also enable diffusion LLMs to scale more effectively to complex, real-world codebases.  

\subsection{Threats to Validity}

\cl{We organize the threats to the validity of our study according to internal validity, external validity, and construct validity.}

\textbf{\cl{Internal Validity.}} \cl{Internal validity concerns whether our experimental procedures support the conclusions drawn about the studied models and settings.} The performance of diffusion LLMs is influenced by multiple factors, which introduces controllability but also poses challenges for replication. To ensure reproducibility, we provide detailed experimental settings and links to the model weights in the supplementary material. In addition, we make our code repository publicly available. \cl{For stochastic decoding settings, we report results across multiple seeds to characterize variability.} These efforts enhance transparency and help ensure that the results of our experiments can be \cl{reproduced consistently} in code generation tasks. \cl{Nevertheless, exact replication of closed-source models such as Mercury-Coder-Small may be affected by API changes or undocumented system updates. Furthermore, some setting analyses are necessarily conducted on a subset of benchmarks because exhaustive sweeps over all models, benchmarks, and decoding settings are expensive; the observed trends may therefore be sensitive to the specific configurations tested, and we restrict our claims accordingly.}

\textbf{\cl{External Validity.}} \cl{External validity concerns the extent to which our findings generalize beyond the evaluated models, benchmarks, and tasks.} \cl{Although our experiments cover function-level synthesis, multilingual generation, recent programming problems, and repository-level understanding, the conclusions are still bounded by the selected benchmarks and model-benchmark configurations. Therefore, we focus our claims on the evaluated settings and avoid treating trends from a single benchmark family as universal properties of diffusion LLMs.} Furthermore, although we have conducted a comprehensive study of 7 diffusion LLMs for code generation, almost all the models we evaluated are smaller than 10B parameters. \cl{Mercury-Coder-Small is closed-source, so its exact scale and training pipeline are unknown.} Recently, larger open-source diffusion LLMs have become available, including the 16B and 100B variants of LLaDA2.0~\cite{bie2025llada20scalingdiffusionlanguage}. Evaluating these larger models and assessing whether our findings generalize across model scales remain directions for future research.

\textbf{\cl{Construct Validity.}} \cl{Construct validity concerns whether our measurements adequately capture the constructs they are intended to represent, namely the code generation effectiveness and decoding efficiency of diffusion LLMs. Following common practice, we operationalize effectiveness using pass@k on established benchmarks and efficiency using throughput, FLOPs/token, and average decoding time; these metrics, however, reflect only specific aspects of code quality and computational cost.} \cl{Since LLMs are trained on vast amounts of data, data leakage poses a potential threat to the validity of our experiments, especially for widely used static benchmarks such as HumanEval and MBPP, whose scores may partially reflect memorization rather than generalization. To reduce this concern, we include the recent LiveCodeBench 2410--2504 split alongside the static benchmarks. This does not eliminate contamination, so results on HumanEval and MBPP should be interpreted together with the newer benchmark.}

\section{Conclusion}
This paper conducts \cl{an empirical study} that systematically investigates the performance of diffusion large language models for code generation. Our results \cl{show that current diffusion models provide promising but uneven performance, and can be competitive with representative AR LLMs in several evaluated settings}. \cl{We further characterize how selected inference factors affect effectiveness and efficiency, providing practical guidance for the evaluated settings. In addition, the three diffusion LLMs show greater robustness than a comparable-scale AR baseline over the evaluated 1k--8k long-context inputs, although this comparison is model-specific and not fully context-window matched.} Building on these findings, we outline several promising directions for future research and highlight opportunities to advance the development of more effective diffusion LLMs for code generation. Overall, our study highlights the \cl{potential} of diffusion LLMs in this domain and calls for greater attention from both academia and industry to advance their application across a wider range of code-related tasks.

\section{Data Availability}
The source code is available at \cl{\url{https://github.com/THU-Agent/dLLM4CodeGeneration}}.

\bibliographystyle{ACM-Reference-Format}
\bibliography{refs}

\end{document}